\documentclass{snam}

\usepackage{txfonts}
\usepackage{amsmath}
\usepackage{graphicx}
\usepackage{sidecap}
\usepackage{subcaption}

\newcommand{\Ec}{\mathcal{E}}
\newcommand{\Sc}{\mathcal{S}}
\newcommand{\Uc}{\mathcal{U}}
\newcommand{\Cc}{\mathcal{C}}
\newcommand{\Tc}{\mathcal{T}}
 
\usepackage{booktabs}
\usepackage[colorlinks=true,citecolor=blue]{hyperref}%
\usepackage[table,xcdraw]{xcolor}
\usepackage{algorithm, algorithmic}
\usepackage{nccmath}
\usepackage{pgfplotstable}

\usepackage[toc,page]{appendix}

\usepackage{enumitem}

\pgfplotsset{compat=1.16}
\begin{document}

	\title{Constrained Expectation-Maximisation for inference of social graphs explaining online user-user interactions}
	\author{Effrosyni Papanastasiou \and Anastasios Giovanidis}

	\institute{Sorbonne University, CNRS, LIP6, F-75005, Paris, France \\ 
	\text{effrosyni.papanastasiou@etu.sorbonne-universite.fr, anastasios.giovanidis@lip6.fr}
	}
	
	\abstract
	{Current network inference algorithms fail to generate graphs with edges that can explain whole sequences of node interactions in a given dataset or \textit{trace}. To quantify how well an inferred graph can explain a trace, we introduce \textit{feasibility}, a novel quality criterion, and suggest that it is linked to the result's accuracy. In addition, we propose CEM-*, a  network inference method that guarantees 100\% feasibility given online social media traces, which is a non-trivial extension of the Expectation-Maximization algorithm developed by Newman (\citeyear{b7}). We propose a set of linear optimization updates that incorporate a set of auxiliary variables and a set of feasibility constraints; the latter takes into consideration all the hidden paths that are possible between users based on their timestamps of interaction and guides the inference toward feasibility. We provide two CEM-* variations, that assume either an Erdős–Rényi (ER) or a Stochastic Block Model (SBM) prior for the underlying graph's unknown distribution. Extensive experiments on one synthetic and one real-world Twitter dataset show that for both priors CEM-* can generate a posterior distribution of graphs that explains the whole trace while being closer to the ground truth. As an additional benefit, the use of the SBM prior infers and clusters users simultaneously during optimization. CEM-* outperforms baseline and state-of-the-art methods in terms of feasibility, run-time, and precision of the inferred graph and communities. Finally, we propose a heuristic to adapt the inference to lower feasibility requirements and show how it can affect the precision of the result.}
	%We believe that feasibility can help us better understand the potential true interactions that exist between nodes but are not observed in the data.
	\maketitle

\section{Introduction}
\label{intro}

Given a set of observed data, \textit{network inference}, or \textit{reconstruction} is the task of determining whether an edge exists or not between any pair of nodes that have interacted at some point in time. Network inference was first used in computational biology, where it was invented as a tool to recreate and explain complex interactions between important nodes, such as proteins or genes (Friedman et al. \citeyear{b11}). Network inference methods have since been applied in a variety of fields besides biology. Examples include epidemiology (Zhang et al. \citeyear{zhangep}; Firestone et al. \citeyear{firestone}), finance (Giesecket et al. \citeyear{giesecke}), and telecommunications (Wu et al. \citeyear{wu}). The main goal of this paper is network inference in the domain of Online Social Networks (OSNs). Their enormous growth in the last decade has resulted in huge amounts of information circulating online from user to user. As a result, research has turned to inference algorithms to derive diverse types of networks which can be useful in various fields such as marketing, advertising, and politics. In advertising, for example, inference algorithms have been employed to derive the probabilities of influence between users, or the way that a specific news piece has diffused on the platform (Gomez-Rodriguez et al. \citeyear{b2d}). 

The reason why non-trivial methods such as network inference are needed to infer these types of networks is lying in the structure of the online datasets themselves. Regarding the diffusion of information through an online platform, the data we can find is limited and does not directly depict how it propagates from user to user. On Twitter, for example, given a tweet by an author and the users that retweeted it, we can get information such as the timestamps of each retweet, but we cannot know where they really retweeted it from\footnote{According to the Twitter API documentation of a Tweet Object, the "retweets of retweets do not show representations of the intermediary retweet, but only the original Tweet." \url{https://developer.twitter.com/en/docs/twitter-api/v1/data-dictionary/object-model/tweet}}. This suggests that inferring the true propagation of a tweet when the friendship graph is unknown is not trivial. Network inference algorithms are thus brought into play and make it possible to infer the real way that information propagates on OSNs by exploiting the available interactions between users (the \textit{trace}). Regarding the learning method itself, different methods have been employed, including maximum likelihood (Harris et al. \citeyear{ml}), expectation-maximization (EM) (Dempster et al. \citeyear{em}), and other models of influence computation, such as Discrete-Time and Continuous-Time Models (Goyal et al. \citeyear{b6}).

When looking at the result of an inference method, one can check whether the input trace is what we call, \textit{feasible}, given the generated network. We can do this by verifying that the inferred graph of connections includes a path from the author of every original post (e.g., tweet) to all other users that shared the post (e.g., via retweets) in the trace. For feasibility, this path should respect the chronological order of the respective interactions in the trace. However, as we will show later experimentally, existing works have disregarded feasibility as a quality criterion that the inferred graph must meet. Therefore, in this paper, we propose trace feasibility as an imperative requirement that must be met by an inference framework applied to OSNs. 

Our intuition behind this proposal is that a feasible graph that can explain all the interactions and their chronological order inside the trace is closer to the real one. This could become more obvious if we think of what non-feasibility entails: suppose that there is an interaction by a user in the trace (e.g., reshare of a post) that cannot be explained by the inferred graph. This means that there is no path (with one or more hops) in the inferred graph from the user author to the user who reshared the post, or that the path is temporally not feasible. Then, either the latter user found this post from some other source (e.g., platform recommendation), or there is an error in the inference because the two users appear disconnected or connected in the wrong (temporally non-feasible) direction. By enforcing feasibility during graph inference, we guarantee that the graph can reproduce and explain all events and interactions observed in the available trace. Of course, in reality, a percentage of the observed interactions can come from indirect diffusion (e.g., recommendations); as we will show later, it is possible to take this into account by assuming some fixed percentage of direct diffusion during the inference process.

Given the above motivation, we can examine whether current methods in the literature infer graphs that guarantee feasibility. By looking into the seminal work of Saito et al. (\citeyear{b5}), we see that the results suffer from the fact that it is not possible to identify the source of influence for a large number of retweets, and therefore their existence in the trace cannot be explained. Therefore, trace feasibility given the inferred network of influence is not achieved. In another fundamental work, Gomez-Rodriguez et al. (\citeyear{b2d}) proposed the NetInf method to infer the optimal network that most accurately explains a sequencing of interactions. However, they only give approximate solutions that, when applied to real-world data, are neither feasible nor accurate. More recently, Newman (\citeyear{b7}) introduced an EM algorithm that is designed for network inference using unreliable data. As the algorithm does not consider that there are hidden paths between the users, the feasibility of the trace given the inferred network is not guaranteed. Building on Newman's work, Peixoto (\citeyear{b12}) was the first to propose a method that performs network reconstruction together with community detection. However, as we will validate experimentally, despite being more precise than the methods above, the results suffer from slow convergence times and again, do not always guarantee feasibility which has an impact on precision. Therefore, as we can see, the inference methods that are currently available in the literature suffer from the fact that they do not explicitly guarantee the feasibility of the results. This is extremely critical since the resulting graphs infer edges that cannot confirm the trace itself. Additionally, as we will show later, each method presents other smaller issues that could have been avoided by enforcing the feasibility guarantees that we propose.

As a solution to the above, we introduce a fresh approach to network inference, which we call CEM-* (Constrained Expectation Maximization). It infers a posterior distribution of feasible underlying graphs that explain the provided social trace while respecting the chronological order of the interactions observed. Since the structure of the underlying graph is not known, the definition of a prior that enforces a structure to the posterior inferred graph is necessary. In this work, we will introduce two special cases of CEM-*: (i) CEM-er, which uses an Erdős–Rényi (ER) prior, and (ii), CEM-sbm, which uses the Stochastic Block Model (SBM). Besides, CEM-* can be adjusted accordingly to include other priors as well. All in all, we enrich the literature with the following contributions:
\begin{itemize}
	\item We define social trace feasibility, and discuss its importance for network inference in the domain of OSNs. To guarantee feasibility, we devise a set of inequalities (constraints) to account for all the possible hidden paths given the timestamps of interaction between the nodes in the social trace (users). \\
	\item We propose CEM-*, a non-trivial extension of the Expectation-Maximization algorithm originally proposed by Newman (\citeyear{b7}) that further incorporates the above set of feasibility constraints. Its main advantage is that it formulates inference as a linear optimization problem, making the task easier to compute. For the graph's unknown distribution, we start with an ER prior, following Newman's (\citeyear{b7}) formulation, and call the method CEM-er (see also our conference version (Papanastasiou \& Giovanidis \citeyear{ours})). \\
	\item We introduce CEM-sbm, a variation of CEM-er that uses an SBM instead of an ER prior that is more realistic to the underlying structure of social graphs. On top of graph inference, CEM-sbm allows us to infer and assign users in communities simultaneously during optimization. Its main benefit against Peixoto (\citeyear{b12}), except for guaranteeing feasibility, is that it is more scalable and easier to compute.\\
	\item We apply CEM-* on a synthetic social trace and compare the inferred graph against the ground truth. We also apply it on a real-world Twitter trace with almost 300,000 tweets and more than 1,600,00 retweets and compare the result against the real friendship graph that we have available. Extensive numerical evaluations of CEM-* against other baseline and state-of-the-art inference methods demonstrate the algorithm's ability to run on large graphs and trace sizes, which is not always guaranteed by the alternatives. \\
	\item We show that real-world traces are not always 100\% feasible given the real graph that underlies them and we propose a technique with which we can tune CEM-* to adapt to lower feasibility requirements. We evaluate to what extent tuning the inferred graph's feasibility can infer edges with better accuracy.
	
\end{itemize}	

The rest of this paper is organized as follows: in Section 2 we present related literature. In Section 3 we introduce the formulation of the problem. Section 4 presents the modeling of the problem and the learning method that we follow. Section 5 describes the datasets that we use and the methodology of the experiments. Sections 6, 7 and 8 show the results of the experiments and the comparison with other methods for the synthetic and the real-world traces respectively. Section 9 presents conclusions and future work. The code for both CEM-er and CEM-sbm is publicly available on GitHub\footnote{\url{https://github.com/effrosyni-papanastasiou/constrained-em}}.

\section{Related literature}
\textit{Graph inference.} Numerous studies have proposed graph inference methods by simultaneously recovering influence probabilities between users. This is usually possible by observing users' infection timestamps from the available cascades of interactions. For example, Goyal et al. (\citeyear{b6}) compute probabilities from a real social graph and a log of actions on Flickr using Continuous and Discrete Time Models with incremental equations. He and Liu (\citeyear{heliu}) presented an approach that recovers a graph from a small number of cascade samples by utilizing the similarities between strongly linked diffusion graphs. A different line of work focuses on learning embeddings to perform the same inference task: for instance, Wang et al. (\citeyear{b2dw}) suggested predicting information diffusion by learning user embeddings that capture unique characteristics both of the diffusion and the network. Later, Bourigault et al. (\citeyear{c2}) presented an embedded version of the IC model on OSNs that learns information diffusion probabilities along with the representation of users in the latent space. (Zhang et al. \citeyear{b2b}) proposed a probabilistic generative model to learn information cascade embeddings that predict the temporal dynamics of social influence.

\textit{Graph inference with incomplete data.} Additionally, many works consider that the observed cascades are incomplete or partially observed, which is frequently the case in real-world settings. This is why a diffusion model must be chosen along with the learning method to represent how we believe that information has been passed through the cascades. Wu et al. (\citeyear{wuetal}) for instance, created an EM method that can tolerate missing observations in a diffusion process that follows the continuous independent cascade (CIC) model. Daneshmand et al. (\citeyear{b2c}) proposed an $L_{1}$-regularized maximum likelihood inference method for a well-known Continuous-Time diffusion model. Lokhov (\citeyear{lokhov}) introduced an approximate gradient descent approach that estimates the influence parameters using gradients of the likelihood calculated via mean-field approximation and dynamic message passing. Their formulation makes the computation tractable, but the complexity of the gradients causes slow convergence.

\textit{Selecting a prior when the ground truth is unknown.} Several link prediction methods extract future or missing links in datasets in which the underlying graph connecting the users is known (Saito et al. \citeyear{b5}, Bourigault et al. \citeyear{c2}, Lagnier et al. \citeyear{b8}, Jin et al. \citeyear{b30}, Peel et al. \citeyear{peel}). However, our goal differs from these types of problems since we have to infer links in a setting where the neighborhoods of the nodes are unknown. We must therefore select a prior structure that is close to the underlying network. For example, Le et al. (\citeyear{b2e}) selected the SBM as the underlying network structure, because of its simplicity and its ability to approximate real networks. Similarly, Peixoto (\citeyear{b12}), used the degree-corrected SBM as a prior, motivated by its ability to inform link prediction when dealing with incomplete or erroneous data. In another example, Newman (\citeyear{b2f}) experimented with different kinds of priors, such as the random graph, the Poisson edge model, and the SBM.

\textit{Neural networks.} In a more recent line of work, recurrent neural networks have been used to predict edges given probability distributions conditioned on temporal sequences of past knowledge graphs (Jin et al. \citeyear{b30}). Neural networks usually require the graph of nodes as input. However, in most social media network settings the friendship graph of user nodes is either not known or has not been published by the creators of the datasets. This makes the use of neural networks for inferring hidden edges more challenging. We leave the use of such methods for network inference when the underlying graph is unknown or incomplete as a future interesting task.
%Some other approaches combine community detection with functional observation. Berthet et al. [31] derived the necessary conditions for the exact recovery of group assignments for dense weighted networks generated with community structure given observed microstates of an Ising model. Hoffmann et al. [32] proposed a way of inferring community structure from time-series data that avoids network reconstruction through direct modeling of group dynamics. However, neither of these approaches attempts to perform network reconstruction together with community detection. Furthermore, they are tied down to one particular inverse problem, and as we will show, our general approach can be easily extended to an open-ended variety of functional models.
\section{Problem formulation}
\label{preprocess}
\subsection{Input data trace}
\begin{figure}[htbp]	\centering{\includegraphics[width=0.7\columnwidth]{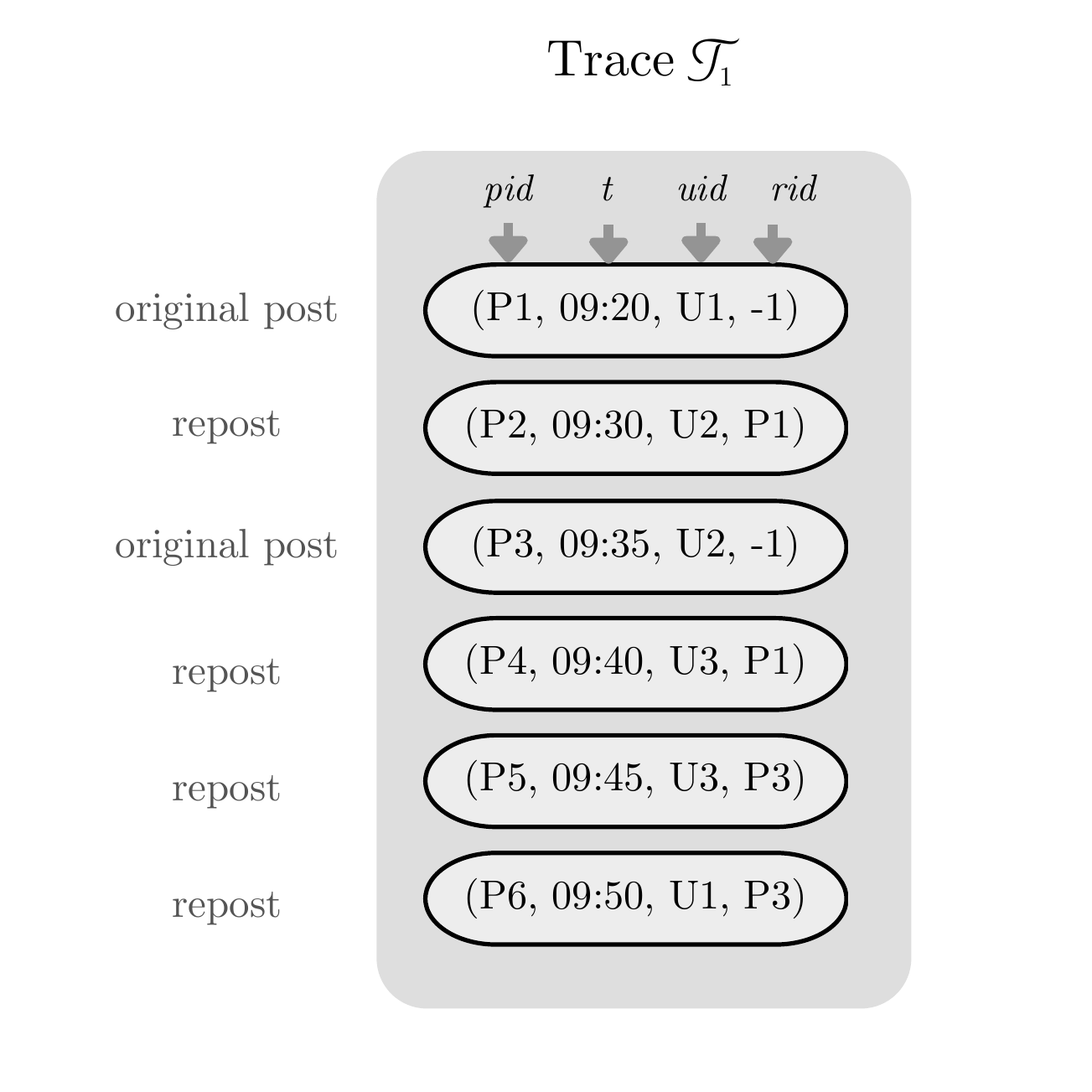}}
	\caption{Example of information available on an OSN trace.}
	\label{fig1}
\end{figure}
As mentioned above, to infer an unknown network we must provide as input a trace of interactions between the nodes of interest. In this paper, as we focus on traces from OSNs, our goal is to infer friendship graphs by looking into the online interactions between users, and more specifically into the \textit{posts} and the \textit{reposts} that they generate. On Twitter, for example, this corresponds to the tweets and retweets that the users exchange.

Throughout the paper, we will use the following notation: the input interaction log with the posts and reposts is denoted by $\Tc$ and it includes $T$ posts/reposts in total. For each instance in the trace, we keep only four types of information: its unique post id ($pid$), the time that the user posted it ($t$), the unique user id ($uid$), and the repost id ($rid$) that equals $-1$ if the post is original, or, if it is a repost, it is equal to some $pid \in \Tc$ which points to the original post instance in the trace. If a user is the author of a $pid$ we mark them as $author_{pid}$. The set that includes all the users that participate in the trace is denoted by $\Uc$ and is of size $|\Uc|=N$. Figure \ref{fig1} shows an example of a trace $\Tc_{1}$ like the one described above. It includes $T=6$ posts/reposts instances and $N=3$ users in total. The first instance in $\Tc_{1}$ is an original post with $pid=$ P1, and is posted at $t=$ 09:20 by author U1; the second instance with $pid= $ P2, tells us that user U2 reposted at $t= $ 09:30 the post with $pid=$ P1 (mapped to the author U1), and so on.

\subsection{Problem formulation} Given an available trace $\Tc$ of the activity of a set of users $\Uc$, we assume that there is an underlying friendship graph $G$ connecting all users in $\Uc$ that is unknown and is what we are trying to infer. More formally, it is a directed friendship graph $G$ where the nodes are the $N$ users in $\Uc$ and each edge $(i,j)$ translates to user $j$ following user $i$. The graph $G$ is represented by an adjacency matrix \textbf{A}, of dimensions $N \times N$, where an element $A_{ij}$ equals $1$ if user $j$ follows user $i$. The current paper aims to infer the hidden adjacency matrix \textbf{A}.
\begin{figure}[htbp]
	\centering{\includegraphics[width=1\columnwidth]{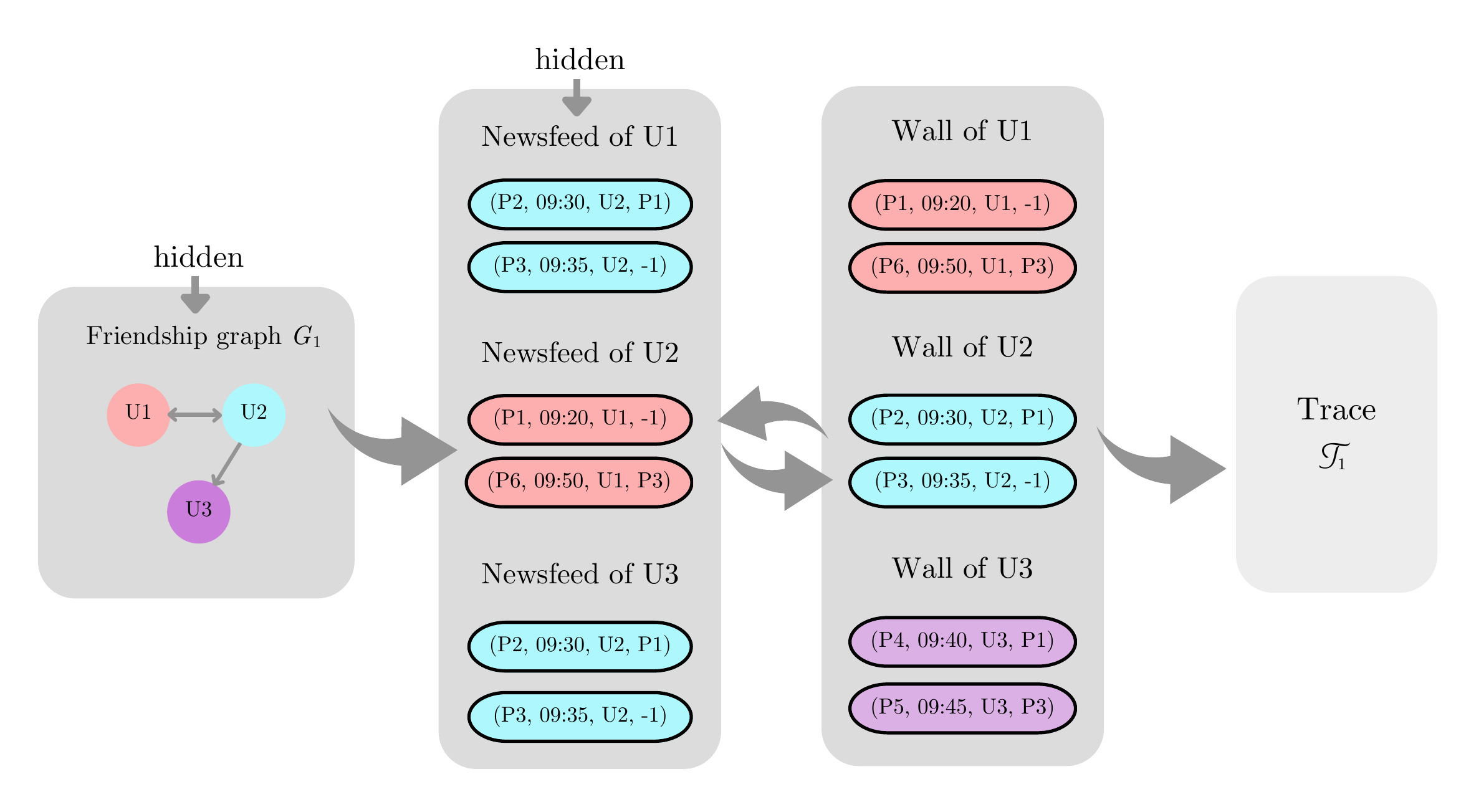}}
	\caption{The hidden way that information diffuses through the ground truth network of users $G_{1}$ that produces the trace $\Tc_{1}$. Our goal is to infer $G_{1}$ (or equivalently its adjacency matrix \textbf{A} from $\Tc_{1}$.}
	\label{fig1b}
\end{figure}

\textbf{Hidden information.} Generally, a social media platform provides a Newsfeed and a Wall for each user. The Wall includes the posts and reposts of the users, whereas the Newsfeed includes the posts and reposts created by their respective followees. Newsfeeds are formed based on the friendships in the network. Accordingly, Fig. \ref{fig1b} shows the possible Newsfeeds and Walls of the users in $\Uc_{1}$ that created the trace $\Tc_{1}$. As we notice, the Newsfeeds are a result of the way that users are connected, i.e., their friendship graph $G_{1}$, which is what we are seeking to infer. Walls are filled with individual posts from users and their interaction with Newsfeeds. If we assume that we have access to the unknown Newsfeeds and the corresponding friendship graph of Fig. \ref{fig1b}, we can infer directly how the post P1 observed in $\Tc_{1}$ is diffused:
\begin{enumerate}
	\item It is initially posted by author U1 at $t_{0}$=09:20.\\
	\vspace{-0.3cm}
	\item At timestamp $t_{0}$ post P1 appears on the Newsfeed of U1's followers, in this case user U2. \\
	\vspace{-0.3cm}
	\item At a later timestamp, $t_{1}$=09:30, U2 reposts P1 on their Wall. Their repost takes the $pid$=P2. \\
	\vspace{-0.3cm}
	\item At the same timestamp $t_{1}$, P2 appears on the Newsfeed of U2's followers, U1 and U3. \\
	\vspace{-0.3cm}
	\item Later, at $t_{2}$=09:40, U3 reposts P2. Their repost takes the $pid$=P4.\\
	\vspace{-0.6cm}
\end{enumerate}
As a result, we inferred that P1 diffused from user U1 to U2 and then to U3 (assuming that users only retweet the users that they follow). Inferring this path was trivial since we assumed that we had access to the Newsfeeds which show the intermediary $pids$ of the reposts. However, until today, social media platforms keep Newsfeeds private to each user.  Therefore, in the final trace $\Tc_{1}$ this information is hidden. Instead, we only have access to the timestamps of the reposts of P1 and the author it is mapped to (user U1). For user U2 it is trivial to infer that they reposted P1 directly from U1 (and thus follow them) since they are the first in the trace to repost it. However, it is non-trivial to infer through whom U3 reposted P1; it could be through any of the users U1 or U2.

Of course, the above example is quite simplistic; we can still come up with some trivial guesses about how the three users are connected that are not very far from the ground truth. In reality, though, we will have to deal with traces that include millions of entries, which makes our task much more challenging. Since social media traces hide the Newsfeeds and the intermediary retweet ids, we do not know the real paths through which posts diffuse: a repost made by each user $uid$ points only to the author of the initial post and not to the real user that $uid$ reposted. Therefore, due to the trace being only a (partial) view of each user's Wall and their interactions with their (hidden) Newsfeed, we cannot infer the friendship connections between the users directly.

Our intuition is that it is more likely that user $j$ is following user $i$ ($A_{ij}=1$) if a post reaches often user $j$ through user $i$ (via the edge ($i,j$)). With this information not being directly available, we aim to infer the intermediary diffusion paths that are hidden in the trace. This will generate the unknown friendship graph $G$ in question. To achieve this, we introduce a set of constraints that guides the graph's inference toward a feasible result. 
%Additionally, as we have assumed that the friendship graph is hidden from us, we cannot infer the paths in a trivial way. We, therefore, claim that friendship graph inference can be facilitated by hidden path reconstruction of posts' paths by leveraging the traces provided by OSNs.
%We expect the task of network inference to be more accurate when the trace includes adequate information between the users.

\subsection{Assumptions on the diffusion of posts}
To generate the hidden diffusion paths, we first need to decide on a diffusion model. In this case, we opt for a simple model, the SI diffusion model, which has been extensively used in epidemiological models (Daley and Gani, \citeyear{b4}) and apply it to social media users: when a new post arrives on a user's Newsfeed, they are Susceptible to infection. If they choose to repost it they become Infected given the specific post and remain so for the rest of the diffusion. Most existing works using the SI model consider that infection can happen only one time step ahead, after a user becomes Susceptible. We assume, however, that when a user posts a message, they can diffuse it to their still uninfected followers (those in the Susceptible state) during any consecutive timestamp. Furthermore, we make some additional assumptions as follows:
\begin{enumerate}
	\item The author of every original post that has been reposted is included in the trace $\Tc$. \\
	\vspace{-0.3cm}
	\item Users repost \textit{only} from their \text{followees}, i.e., the users they follow. We assume that the latter are always present in the available trace.\\
	\vspace{-0.3cm}
	\item A post can diffuse from user $i$ to user $j$ only if user $i$ has shared the post chronologically earlier in $\Tc$ than user $j$.
\end{enumerate}
Although the second assumption does not always hold in practice, it simplifies our task. As we will see later, our approach can be expanded accordingly to take into account instances in which people repost content from followees who are not inside the trace or even from users outside their list of followees (e.g., when Twitter users repost something from the trending hashtags or via the search function, etc). We should also note that we can only obtain friendships between users who have interacted with one another at least once in the available trace $\Tc$.
%In the journal version we also have introduced the variation with SBM prior to achieve better inference by searching the graph structure for communities. 

\subsection{Episodes}
We collect the set of all original posts (the ones that have $rid=-1$ in $\Tc$) that we call $\Sc$. Each original post $s \in \Sc$ along with its reposts is called an \textit{episode} and is defined as follows: \begin{definition}[Episode]
	For each original post $s \in \Sc$ we define an {episode} as a set of users $\Ec_{s}$ = $author_{ s} \cup \{u \in \Uc \text{ }| \text{ } \exists \text{ } (pid,t): (pid, t, u, s) \in \Tc \}$. In other words, each episode $\Ec_{s}$ includes the author of $s$, denoted by $author_{ s}$, followed by the users who reposted it, in chronological order. The whole set of episodes is denoted by $\Ec$ and includes $S$ episodes in total.
\label{episodefeas}
\end{definition}To indicate that user $i$ appears in $\Ec_{s}$ before $j$ we use the notation $  i <^{s} j$. We call this pair a \textit{temporally ordered pair $(i,j)_{s}$}. Out of the $S$ total episodes in $\Tc$, we count $M_{ij}$ where it holds that $\small i <^{s} j$. If $M_{ij} > 0$, it is probable that $j$ has reposted content from $i$. In this case, the pair is referred to as an \textit{active pair}. Our intuition is that we become more certain about the existence of a diffusion path from $i$ to $j$ as $M_{ij}$ becomes larger. As a result, $M_{ij}$ is a quantity that can determine the hidden post-propagation paths and we will use it extensively in the sections that follow. Every piece of information that can be directly derived from a trace $\Tc$ can be found in Table \ref{tab1}.

\begin{table}[htbp]
	\caption{Information that is directly available from the data.}
	\resizebox{1\linewidth}{!}{
		\begin{tabular}{ll}
			\hline
			\addlinespace[0.1cm]
			\text{Symbol} & {Definition}\\
			\addlinespace[0.1cm]
			\hline
			\addlinespace[0.1cm]
			$\Tc$ & Set of $T$ post instances of the type $(pid, t, uid, rid)$.\\
			$\Uc$ & Set of users that are included in $\Tc$ ($|\Uc|=N$).\\
			$\Sc$ & Set of original posts in $\Tc$ ($|\Sc|=S$).\\
			$\Ec$ & Set of episodes in $\Tc$ ($|\Ec|=S$). \\
			$\Ec_{s} \in \Ec$ &Episode of original post $s$, $1 \leq s \leq S$. \\
			$author_{ s}$ & The $uid$ of the author of $s$. \\
			\small $ i <^{s} j$ & User $i$ reposted or posted $s$ before user $j$. \\
			$M_{ij}$ &\# episodes where it holds true that $i <^{s} j$. \\
			%$\Cc$ & Set of constraints $\Cc = \{c_{1}, c_{2}, ..., c_{({P-S}}\}$ \\
			\addlinespace[0.1cm]
			\hline
	\end{tabular}}
	\label{tab1}
\end{table}
\subsection{Feasibility of a trace given an inferred graph}
\label{feas}
For every episode $\Ec_{s}$ in the trace $\Tc$ and every user $i$ that reposted $s$ before $j$ in time, we define the binary variable $X_{ij}(s) \in \{0,1\}$ that is equal to $1$ if the post $s$ passed from $i$ to $j$ (i.e., $j$ follows $i$) and $0$ otherwise. As underlined in the previous section, the real value of $X_{ij}(s)$ is unknown. Therefore, given the chronological order of reposts in $\Ec_{s}$, we may imagine many feasible routes through which the post $s$ might have spread to those who reposted it. These paths create a {propagation graph} $G_{s}=\{V_{s}, E_{s}\}$ per episode, with the users in each episode $\Ec_{s}$ as nodes ($V_{s}=\Ec_{s}$), and the edges set $E_{s}$ containing the (unknown) edges that we infer for the given post. Every edge that we infer follows the propagation's direction; for instance, an edge $(i,j)$ inferred in $G_{s}$ indicates that $X_{ij}(s)=1$. Given the above and our problem definition, for each episode $s$ in $\Tc$, our goal is to infer a directed acyclic graph (DAG) $G_{s}$ that is \textit{feasible} and explains the whole $\Ec_{s}$ sequence.
\begin{definition}[Feasible propagation DAG $\boldsymbol G_{s}$ per episode  $\boldsymbol \Ec_{s}$] \label{def2}
	Given an episode $\Ec_{s}$ from $\Tc$, we say that a propagation DAG $G_{s}$ is feasible, or, equivalently, that it explains $\Ec_{s}$, if (i) there exists (at least) one directed path from the author $author_{ s}$ to every other user $j \in \Ec_{s} \backslash author_{ s}$ and (ii), for each edge $(i,j)$ of the path it holds that $i <^{s} j$, i.e., all of its edges follow the time-ordering of the reposts. If we take the union of every feasible propagation graph $G_{s}$ inferred per episode $\Ec_{s}$, we get the full friendship graph $G$ and we can build its adjacency matrix \textbf{A} as follows: we set $A_{ij}=1$ if there exists at least one $G_{s}$ where the edge $(i,j)$ exists, and $0$ otherwise.
\end{definition}
\begin{definition}[Feasible friendship graph $\boldsymbol G$]
	An inferred graph $G$ is called {feasible}, if, for every episode $\Ec_{s}$ in $\Tc$, there exists a subgraph which is a feasible propagation DAG $G_{s}$ as we defined it above. Keep in mind that the full graph $G$ is not a DAG.
\end{definition}
\begin{figure*}[htbp!]
	\centering
	{\includegraphics[width=0.8\textwidth]{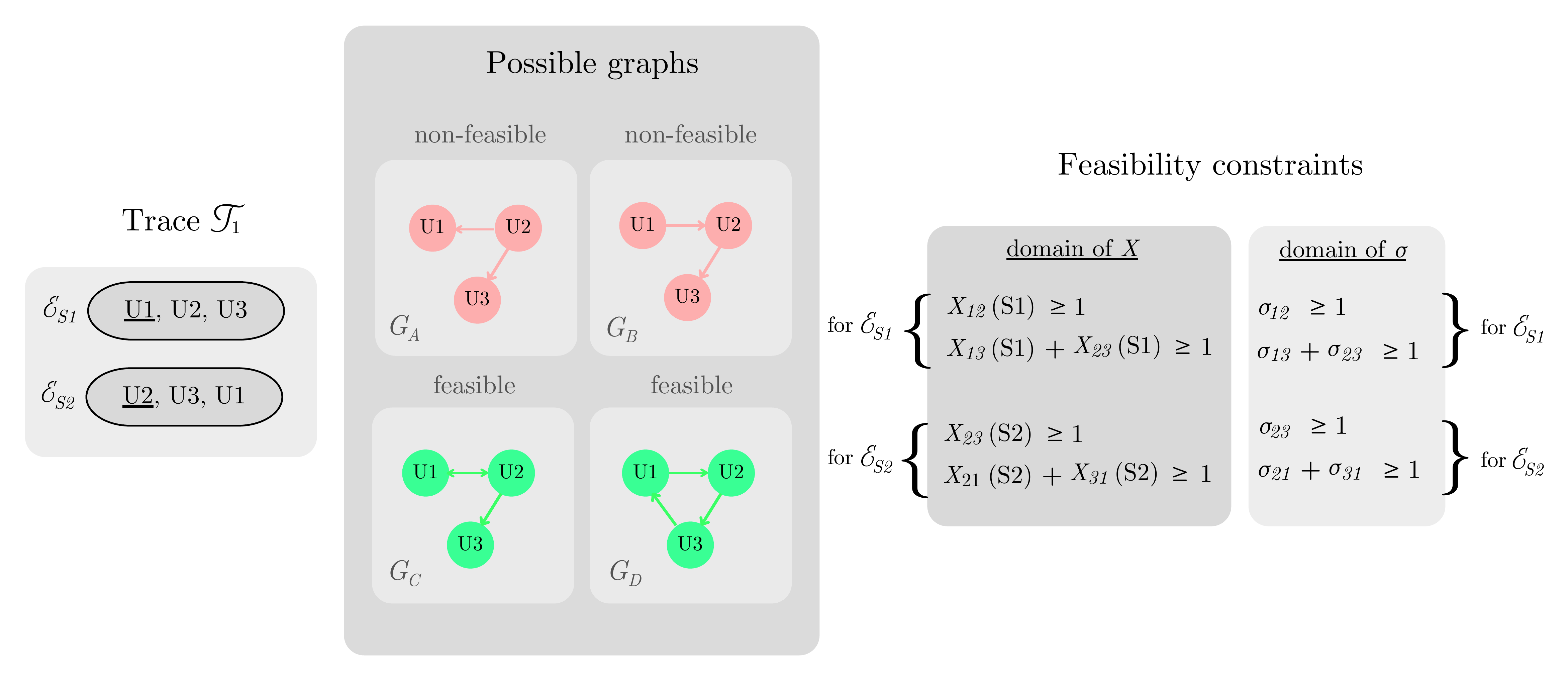}}
	\caption{Feasibility check of different inferred graphs given a trace $\Ec=\{\Ec_{\textit{S1}},\Ec_{\textit{S2}}\}$ .} \label{fig2}
\end{figure*}
To make the concept of feasibility more clear, we show in Fig. \ref{fig2} some examples of possible friendship graphs that could have been inferred, given the example trace $\Tc_{1}$. It contains two episodes $\Ec=\{\Ec_{\textit{S1}},\Ec_{\textit{S2}}\}$, where S1 = P1 and S2 = P3. Graph $G_{A}$ explains episode $\Ec_{\textit{S2}}$ by inferring that the post S1 diffused directly from author U2 to users U3 and U1. However, for post S1, there is no feasible path from the author U1, to users U2 and U3 that reposted it. Thus, the episode $\Ec_{\textit{S1}}$ is non-feasible given $G_{A}$ and the final friendship graph $G_{A}$ is only 50\% feasible, since it only explains half the trace. Similarly, graph $G_{B}$ is only 50\% feasible since it does not explain episode $\Ec_{\textit{S2}}$: it does not give a feasible propagation path to explain how S2 arrived to U1 from author U2. In contrast, graphs $G_{C}$ and $G_{D}$ are both 100\% feasible because we can find a feasible propagation graph for each episode in the trace. Therefore, either of the two graphs could be considered a feasible solution to our graph inference problem. We should note here that there are more combinations of feasible connections that we could think of; these figures demonstrate only two representative feasible examples.
\subsection{Inference of post diffusion}
\subsubsection{Feasibility constraints on reposting behavior}
\label{sigma}
The main challenge of network inference in OSNs arises from the fact that the binary value $X_{ij}(s)$ defined in Section \ref{feas} for the different user pairs is unknown. However, we can restrict the number of solutions by imposing a set of constraints on all the values. These constraints should ensure that all the episodes in the trace are feasible given the inferred graph according to Definition 1. Specifically, they should guarantee that if a user $j$ appears in an episode $\Ec_{s}$ (after the author $author_{ s}$) they should be connected with at least one user $i$ that appears in $\Ec_{s}$ before them, including the author of $s$ (i.e., it should hold that $i <^{s} j$). As a result, the constraints have the following format:
\begin{equation}
\sum_{i \in \Ec_{s} \text{ s.t. } i <^{s} j}{X}_{ij}(s) \geq 1, \forall j \in \Ec_{s} \backslash \{author_{ s}\},
\label{eq1}
\end{equation}
\vspace{-0.5cm}
\begin{equation}
\text{ } \text{ } \text{ } \text{ } {X}_{ij}(s) \in \{0,1\}, \text{ } \forall i,j \in \Uc, \text{ } \forall s \in \Sc.
\label{eq1b}
\end{equation}
Fig. \ref{fig2} shows the constraints on $X_{ij}(s)$, given the set of episodes $\Ec=\{\Ec_{\textit{S1}},\Ec_{\textit{S2}}\}$. The role of these constraints is to guide the process toward solutions that belong to the feasible group of graphs. To do so, the constraints should be defined for each episode $\Ec_{s} \in \Ec$, and each user that reposted $s$, according to Eq. \ref{eq1}. For example,  as we see in Fig. \ref{fig2}, given the first constraint for episode $\Ec_{\textit{S1}}$, we can derive easily that the user U2 reposted post S1 directly from its author U1 ($X_{12} (\text{S1})=1$). The second constraint tells us that user U3 has reposted S1 either from U1, or from U2 (or, from both). If we look closer, the possible graphs that we marked as non-feasible earlier violate these constraints. For example, $G_{A}$ violates the first constraint for $\Ec_{\textit{S1}}$, since $X_{12}(\text{S1})=0$. Likewise, $G_{B}$ violates the last constraint for $\Ec_{\textit{S1}}$, since $X_{21}(\text{S2}) + X_{31}(\text{S2}) = 0$. As we saw in the figure and the equations above, the $X_{ij}$ value of a pair $(i,j)$ is different for each episode that it appears in. For example, $X_{23}$ appeared two times, one time for $S1$ and another one for S2. 

With all the possible combinations that each $X_{ij}$ value can take for all active pairs and episodes observed, we soon realize that the problem is intractable when dealing with large traces. The only direct knowledge we have for each pair is the constant value $M_{ij}$, i.e., the total number of times that a user $i$ appears before $j$ in every episode $\Ec_{s} \in \Tc$. What we are interested in, is the number of times that a post diffused through the edge $(i,j)$, out of the $M_{ij}$ times that it could be possible. We model this with the unknown quantity $Y_{ij}$ which is equal to the total number of times that $j$ reposts from $i$. More formally:
\begin{equation}
Y_{ij} = \sum_{s \in \Sc, \text{ s.t. } i <^{s} j}^{\Sc} X_{ij}(s).
\label{eq2}
\end{equation}
As we can see above, to find $Y_{ij}$, we sum over all episodes where it holds that $i <^{s} j$. This happens $M_{ij}$ times in total.

\textbf{Diffusion probabilities.} To solve the problem we make the following important assumption: for every active pair $(i, j) $ in any episode $\Ec_{s} \in \Ec$, a user $j$ reposts an $s$ from $i$ independently from other episodes with an unknown diffusion probability $\sigma_{ij} \in [0,1]$. Therefore, $X_{ij}(s)$ is an independent Bernoulli random variable with a mean parameter $\sigma_{ij}$ which does not depend on $s$. In other words, the diffusion probability $\sigma_{ij}$ of a user pair is the same across all episodes, which means that there is no preference in terms of content when someone chooses to repost. Of course, this does not accurately reflect reality but it serves as a useful simplification. Therefore, for an ordered user pair $(i,j)_{s}$, $\sigma_{ij}$ equals:
\begin{equation}
\sigma_{ij} = \mathbb{E}\left[X_{ij}(s)\right].
\label{eq3}
\end{equation}
We can now transfer our problem from searching over the binary domain of $X_{ij}(s)$ to solving over the real domain of the $\sigma_{ij}$ values. By taking the expectation in \eqref{eq1} and given Eq. \ref{eq3}, we get the following set of constraints:
\begin{equation}
\sum_{i \in \Ec_{s} \text{ s.t. } i <^{s} j}\sigma_{ij} \geq 1, \forall j \in \Ec_{s} \backslash \{author_{ s}\},
\label{eq4a}
\end{equation}
\vspace{-0.55cm}
\begin{equation}
\text{ } \text{ } \text{ } \text{ } \sigma_{ij} \in [0,1], \text{ } \forall i,j \in \Uc.
\label{eq4}
\end{equation}
From Eq. \ref{eq2} and Eq. \ref{eq3}, $Y_{ij}$ is the sum of $M_{ij}$ independent Bernoulli random variables that have a mean value $\sigma_{ij}$. In other words, $Y_{ij}$ is an independent Binomial random variable with mean value $M_{ij}\sigma_{ij}$:
\begin{equation}
\mathbb{E}[Y_{ij}] = \sum_{s \in \Sc, \text{ s.t. } i <^{s} j}^{\Sc} \mathbb{E}[X_{ij}(s)] = \sum_{s \in \Sc, \text{ s.t. } i <^{s} j}^{\Sc} \sigma_{ij} = M_{ij}\sigma_{ij}.
\label{eq7d}
\end{equation}
% \subsubsection{Removing redundant constraints}
% \label{rem-con}
% We notice that given a trace, some constraints become redundant and can be removed according to the following rules:
% \begin{itemize}
% \item If all parameters $\sigma_{ij}$ that are included in a constraint $c_{k} \in \Cc$, are also included in a different constraint $c_{w} \in \Cc$, then $c_{w}$ is removed from $\Cc$.
% \item In \eqref{eq4a}, we observe that the first constraint of each episode includes only one variable, which is the $\sigma_{ij}$ between the first user $i=author_{ s}$ and the second user $j$ in the episode. Therefore, given also that $\sigma_{ij} \in [0,1]$, all parameters between the first and the second user of each episode become $\sigma_{ij}=1$. As a result, the first constraint per episode is removed, since the solution for these parameters has already been found.
% \end{itemize}
%Generally, the exact number of constraints by which our problem will be reduced depends on the characteristics of each trace.
\section{Problem Modeling and Learning Method}\label{modeling}
We introduce a feasible inference method, called CEM-*, with two special cases, depending on the assumed distribution of the underlying graph. The first case assumes an Erdős–Rényi (ER) prior and is called CEM-er. According to this prior, the underlying graph that we are trying to infer has been created under a uniform probability $\rho$ that is the same for all edges. However, this does not accurately reflect the structure of social media graphs, which are less random and have some important properties, such as the existence of hubs. After this section, we propose an additional case that incorporates a more realistic model for the underlying graph, the stochastic block model (SBM). We call this extended method CEM-sbm.
%After presenting CEM-sbm, we consider it useful to present again the basic equations involved in CEM-er to (i) understand better the difference between the two priors and (ii), translate more wisely the results of the comparison between our method and other existing methods from the literature.
%We invite the readers to refer to the cited paper \cite{} for more details wherever needed.
%\begin{table}[]
%\centering
%\caption{The problem's unknown variables.}
%\begin{tabular}{ll}
% \hline
% \addlinespace[0.1cm]
% \text{Variable} & {Definition} \\ \hline
% \addlinespace[0.1cm]
% $X_{ij}(s) \in \{0,1\}$ & $X_{ij}(s) = 1$ if user $j$ reposted $s$ by user $i$ \\
% \addlinespace[0.1cm]
% $Y_{ij} = \sum_{m=1}^{M_{ij}}X_{ij}(s) $ & \#times in total that user $j$ reposted $i$ \\
% \addlinespace[0.1cm]
% $\sigma_{ij} = \mathbb{E}\left[X_{ij}(s)\right] $& Probability that user $j$ reposted from $i$\\
% \hline
%\end{tabular}
%\end{table}
\subsection{Erdős–Rényi prior (CEM-er)}
As mentioned above, the prior structure of the network \textbf{A} is not known, and therefore a uniform prior $\rho$ is assumed for all edges. Hence, the prior takes the form of a probability distribution $P(\textbf{A} \text{ } | \text{ } \theta)$, where $\theta$ is a set of hidden parameters that give us more details on the underlying network. Given a trace $\Tc$ of posts and reposts, ${P}(\textbf{A}, \theta \text{ }| \text{ }\Tc)$ is the probability that the inferred graph is \textbf{A} and the parameters get the value $\theta$. The parameters $\theta$ should account for a wider range of potential graph types and data generation methods. Therefore, they are chosen as follows:
\begin{itemize}
	\item The probability that a user $j$ shares content through a user $i$, represented by the set of $\sigma_{ij}$ values that we presented in Section \ref{sigma}.
	\vspace{0.1cm}
	\item To model the uncertainty about the structure of the graph's adjacency matrix \textbf{A}, we assumed that there is a prior probability $\rho$ of an edge drawn independently between any two nodes $i,j$ (Erdős–Rényi prior).
	\vspace{0.1cm}
	\item The \textit{true positive utilization rate $\alpha$}: the probability of a post propagating
	through an edge that we inferred to exist in the underlying graph $G$. Given the (hidden) number of interactions between users $Y_{ij}$, we consider that when an edge exists in $G$ ($A_{ij}=1$) the $Y_{ij}$ out of the $M_{ij}$ experiments are successful (we get $Y_{ij}$ true positive edges in total), each with probability $\alpha$.
	\vspace{0.1cm}
	\item The \textit{false positive utilization rate $\beta$}: the probability of inferring that a post propagated through edges that do not exist in $G$. Likewise to above, when $A_{ij}=0$, we consider that the $Y_{ij}$ out of $M_{ij}$ experiments are successful (we get $Y_{ij}$ false positive edges), each with probability $\beta$.
\end{itemize}
We can see that the global parameters $\alpha$ and $\beta$ depend on whether an edge exists in the ground truth graph $G$. To find the most probable value of the parameters $\theta$ given the observed data and infer $\textbf{A}$ with maximum likelihood, we will employ an Expectation–Maximization (EM) algorithm which is a standard inference tool when some data is unknown or hidden. As suggested by its name, an EM iteration involves two consecutive steps: an expectation (E) step, which computes the expected log-likelihood under the most recent estimation of the parameters in $\theta$; then, a maximization (M) step, which determines the parameters that maximize the expectation. Then, the computed parameters are used in the following iteration, and so on, until we satisfy a convergence criterion.

We start constructing the EM iterations, following the method proposed by Newman (\citeyear{b7}) and employ the Bayes' theorem:
\begin{equation}
{P}(\textbf{A},\theta \text{ }|\text{ } \Tc) = \frac{{P}(\Tc \text{ }|\text{ }\textbf{A}, \theta){P}(\textbf{A}\text{ }|\text{ }\theta){P}(\theta)}{{P}(\Tc)}.
\label{eq5a}
\end{equation}
The probability that we get the specific set of posts and reposts $\Tc$ given \textbf{A} and the parameters $\theta=$\{$\alpha, \beta, \rho, \boldsymbol{\sigma}$\}, found in the numerator of the above expression, will differ here from Newman since we have introduced the hidden number of interactions between users, $Y_{ij}$. Given the ordered nodes of an episode, each repost path is chosen independently per episode. We also assumed as prior knowledge that between any two
nodes in \textbf{A} an edge has been drawn with probability $\rho$. Therefore we get:
\begin{align}
{P}(\Tc \text{ }|\text{ }\textbf{A}, \theta){P}(\textbf{A}\text{ }|\text{ }\theta) = \prod_{i \neq j}{\left[ \alpha^{Y_{ij}}{(1 - \alpha)}^{M_{ij} - Y_{ij}}\rho\right]}^{A_{ij}} \ \nonumber
\\
{ \left[\beta^{Y_{ij}}{(1 - \beta)}^{M_{ij} - Y_{ij}}(1-\rho)\right]}^{1-A_{ij}}.
\label{eq8a}
\end{align}
Given this type of model, when $A_{ij}=1$, the $Y_{ij}$ out of the $M_{ij}$ experiments are successful, each with probability $\alpha$. When $A_{ij}=0$, the $Y_{ij}$ out of $M_{ij}$ experiments are successful, each with probability $\beta$. For the whole set of parameters $\theta$, we assume a uniform prior probability ${P}(\theta)$. If we sum \eqref{eq5a} over all possible networks \textbf{A}, we find that ${P}(\theta \text{ }|\text{ } \Tc) = \sum_{\textbf{A}} {P}(\textbf{A}, \theta \text{ }|\text{ } \Tc)$. Then, as suggested by Newman (\citeyear{b7}), we can apply the well-known Jensen's inequality on the $\log$ of $P(\theta \text{ }|\text{ } \Tc)$:
\begin{equation}
\log {P}(\theta \text{ }|\text{ } \Tc) = \log \sum_{\textbf{A}} {P}(\textbf{A}, \theta\text{ }|\text{ } \Tc) \geq \sum_{\textbf{A}} q(\textbf{A}) \log \frac{{P}(\textbf{A}, \theta\text{ }|\text{ } \Tc)}{q(\textbf{A})},
\label{eq7aa}
\end{equation}
where $q(\textbf{A})$ is any probability distribution over networks \textbf{A} satisfying $\sum_{\textbf{A}}q(\textbf{A}) = 1$. We also define the posterior probability of an edge existing between $i$ and $j$ by $Q_{ij} = {P}(A_{ij} = 1| \Tc, \theta) = \sum_{\textbf{A}}q(\textbf{A})A_{ij}$. If we take the expectation of Eq. \eqref{eq7aa} we find that:
\begin{align}
\label{eq7fer}
\mathbb{E}[\log{P}(\theta \text{ }|\text{ } \Tc)] \geq \sum_{\textbf{A}} q(\textbf{A})\log\frac{D_{ij}}{q(\textbf{A})},
\end{align}
\vspace{-0.3cm}
\begin{multline} \label{eq7fba}
\text{where } D_{ij} = \Gamma \prod_{i \neq j}{\left[ \rho \alpha^{M_{ij}\sigma_{ij}}{(1 - \alpha)}^{M_{ij}(1-\sigma_{ij})}\right]}^{A_{ij}} \\ {\left[(1-\rho)\beta^{M_{ij}\sigma_{ij}}{(1 - \beta)}^{M_{ij}(1-\sigma_{ij})}\right]}^{1-A_{ij} }.
\end{multline}
We find that the choice of $q$ that achieves equality of \eqref{eq7fer} and hence, maximizes the right-hand side with respect to $q$ is:
\begin{equation}
q(\textbf{A}) = \prod_{i \neq j}Q_{ij}^{A_{ij}}(1-Q_{ij})^{1-A_{ij}},
\label{eq22er}
\end{equation}
where, $Q_{ij}$ is the posterior probability that the edge ($i$, $j$) exists, and we find that it equals:
\small
\begin{equation}
{Q_{ij}=\dfrac{\rho \alpha^{M_{ij}\sigma_{ij}} (1-\alpha)^{M_{ij}(1-\sigma_{ij})}}{\rho \alpha^{M_{ij}\sigma_{ij}} (1-\alpha)^{M_{ij}(1-\sigma_{ij})} + (1-\rho) \beta^{M_{ij}\sigma_{ij}} (1-\beta)^{M_{ij}(1-\sigma_{ij})}}}.
\label{eq23ER}
\end{equation}
\normalsize
The details of the above derivation are shown in Appendix A. Hence, to find the maximizing posterior distribution $q(\textbf{A})$ it suffices to find the individual maximizing posterior probabilities $Q_{ij}$ according to Eq. \eqref{eq23ER}. Given these values, if we further maximize with respect to the parameters $\theta=$\{$\alpha, \beta, \rho, \boldsymbol{\sigma}$\} we can get the maximum-likelihood value we seek. The updates for the first three parameters are thus calculated to be the following:
\begin{equation}
\label{ab}
\alpha = \dfrac{\sum_{i\neq j} M_{ij} \sigma_{ij}Q_{ij}}{\sum_{i\neq j} M_{ij} Q_{ij}}, \text{ }\beta = \dfrac{\sum_{i\neq j}M_{ij}\sigma_{ij}(1-Q_{ij})}{\sum_{i\neq j}M_{ij}(1-Q_{ij})}, \text{ }
\end{equation}
\begin{equation}
\rho = \dfrac{1}{{N(N}-1)} {\sum_{i\neq j} Q_{ij}} \label{eq21},
\end{equation} \normalsize where $N$ is the number of users in the trace. Finally, to find the whole vector $\boldsymbol{\sigma}$ that includes all the $\sigma_{ij}$ unknown diffusion parameters, we must solve a linear optimization problem as follows (for derivation refer to Appendix A):
\normalsize
\begin{equation}
\max_{\sigma} \sum_{i \neq j}\sigma_{ij}(W_{ij} - \lambda c)
\label{eq17a}
\end{equation}
\vspace{-0.6cm}
\begin{equation}
\text{s.t. } \boldsymbol{\sigma} \in {F}_{\boldsymbol{\sigma}},	 \nonumber
\end{equation}
\vspace{-0.7cm}
\begin{equation}
\text{where } W_{ij} = M_{ij}\left(Q_{ij}\log\frac{\alpha}{1-\alpha} +(1-Q_{ij})\log\frac{\beta}{1-\beta}\right), \nonumber \\
\end{equation}
\vspace{-0.5cm}
\begin{equation}
\lambda > 0 \text{ some given penalty for regularisation, and $c = \max_{(i,j) \in W} W_{ij}$. \nonumber}
\end{equation}
We added the value $\lambda$ into the optimization objective as a penalty per iteration, since our initial goal is to infer a graph that is feasible with the minimum possible number of edges. Without it, all $(i,j)$ pairs with $W_{ij} > 0$ would immediately get their $\sigma_{ij}=1$, leading to the inference of more edges than we initially wanted. As $\lambda$ moves closer to 1, it forces the optimization goal to be negative and thus, to be guided only by the provided constraints. It is equivalent to penalizing the total expected number of inferred edges. As $\lambda$ approaches 0, the optimization infers the largest number of edges possible. We will explore in detail the effect of the hyperparameter $\lambda$ with values that vary from 0 to 1 in the Experiments section. The final CEM-er algorithm is shown in Algorithm \ref{algo1}.
\subsection{Stochastic block model prior (CEM-sbm)}
%We start by defining the model which best represents our past knowledge of the structure of the underlying network $\textbf{A}$.
Since we are working with social media data, where there is usually a strong presence of communities, we believe it is more realistic to assume that the network is derived from a stochastic block model (SBM), a generative model of community structure that was first proposed in the 1980s by Holland et al. (\citeyear{holland}). In the standard SBM, each node $i$ participates in a different block (community) which we indicate by $g_{i}$, where $i$ may take values in $[1,G]$ where $G$ is the number of hidden communities. The number of edges between nodes $i$ and $j$ follows a Bernoulli distribution with mean $\omega_{g_{i},g_{j}}$, that is the relative probability of intra-community (if $g_{i}=g_{j}$) or inter-community (if $g_{i} \neq g_{j}$) connection.

As we can see, in the case of CEM-er, the prior structure of the network \textbf{A} was the only kind of unobserved data, but in this case, we have two unknowns: the network \textbf{A} and the vector of the group assignments of the users \textbf{g}. Hence, the prior takes the form of a probability distribution $P(\textbf{A}, \textbf{g} \text{ } | \text{ } \theta)$, where $\theta$ denotes the unknown parameters of the distribution, which gives additionally the details of the community structure. This approach, therefore, allows us to infer both the unknown network structure and the community structure simultaneously.
Given a trace $\Tc$, ${P}(\textbf{A}, \textbf{g}, \theta \text{ }| \text{ }\Tc)$ is the probability that we get \textbf{A}, the users' community participation vector \textbf{g} and a set of chosen parameters $\theta$. The parameters set $\theta$ that we select here includes two newly added parameters that replace the prior $\rho$ that we had in the CEM-er case:
\begin{itemize}
	\item Following the SBM for \textbf{A} and the users' community participation vector $\textbf{g}$, we suppose that there is a prior probability $p$ of an edge existing between any two nodes $i,j$ that belong in the same community, i.e., $g_{i} = g_{j}$.
	\item The nodes that belong in different communities are connected with a probability $q$.
\end{itemize}
We construct the EM iterations as we did before, following Bayes' theorem:
\begin{equation}
{P}(\textbf{A}, \textbf{g}, \theta \text{ }| \text{ }\Tc) = \frac{{P}(\Tc\text{ }| \text{ }\textbf{A}, \textbf{g}, \theta){P}(\textbf{A}, \textbf{g}\text{ }| \text{ }\theta){P}(\theta)}{{P}(\Tc)}.
\label{eq5}
\end{equation}
Taking into consideration the definition of the parameters above, the probability that we get the specific trace $\Tc$, given \textbf{A}, \textbf{g} and $\theta=$\{$\alpha, \beta, p, q, \boldsymbol{\sigma}$\} is driven by the probabilities $\alpha$ and $\beta$, whereas the probability that we get \textbf{A} and \textbf{g} given $\theta$ depends on the probabilities $p$ and $q$. Therefore, assuming that each user reposts independently from others:
\begin{equation}
{P}(\Tc\text{ }| \text{ }\textbf{A}, \textbf{g}, \theta){P}(\textbf{A}, \textbf{g}\text{ }| \text{ }\theta) = \prod_{\substack{{i \neq j}\\ {g_{i} = g{j}}}}{\left[ \alpha^{Y_{ij}}{(1 - \alpha)}^{M_{ij} - Y_{ij}}p\right]}^{A_{ij}} \ \nonumber
\end{equation}
\vspace{-0.5cm}
\begin{equation}
 {\left[\beta^{Y_{ij}}{(1 - \beta)}^{M_{ij} - Y_{ij}}(1-p)\right]}^{1-A_{ij}} \prod_{\substack{{i \neq j}\\ {g_{i} \neq g{j}}}}{\left[ \alpha^{Y_{ij}}{(1 - \alpha)}^{M_{ij} - Y_{ij}}q\right]}^{A_{ij}} \nonumber
\end{equation}
\vspace{-0.7cm}
\begin{equation}
{\left[\beta^{Y_{ij}}{(1 - \beta)}^{M_{ij} - Y_{ij}}(1-q)\right]}^{1-A_{ij}}. \label{eq8}
\end{equation}
\normalsize
For the whole set of parameters $\theta$, we assume again a uniform prior probability ${P}(\theta)$. We sum \eqref{eq5} over all possible networks \textbf{A} and we find that ${P}(\theta \text{ }| \text{ } \Tc) = \sum_{\textbf{A}} {P}(\textbf{A}, \textbf{g}, \theta \text{ } | \text{ } \Tc)$. Then, we can apply the well-known Jensen's inequality on the $\log$ of ${P}(\theta\text{ }| \text{ }\Tc)$:
\small
\begin{equation}
\log {P}(\theta\text{ }| \text{ }\Tc) = \log \sum_{\textbf{A}} {P}(\textbf{A}, \textbf{g}, \theta\text{ }| \text{ }\Tc) \geq \sum_{\textbf{A}} q(\textbf{A}, \textbf{g}) \log \frac{{P}(\textbf{A}, \textbf{g}, \theta\text{ }| \text{ }\Tc)}{q(\textbf{A}, \textbf{g})},
\label{eq7a}
\end{equation}
\normalsize
where $q(\textbf{A}, \textbf{g})$ is any joint probability distribution over networks \textbf{A} and group assignments \textbf{g} satisfying $\sum_{\textbf{A}}q(\textbf{A}, \textbf{g}) = 1$. We also define the posterior probability of an edge existing between $i$ and $j$ that belong to communities $g_{i},g_{j}$ by $Q_{ij}(g_{i},g_{j}) ={P}(A_{ij} = 1 \text{ }| \text{ } \Tc, \theta) = \sum_{\textbf{A}}q(\textbf{A}, \textbf{g})A_{ij}$.

For the E-step of the EM algorithm, following the same derivation logic as in the CEM-er variation (in detail in Appendix B), we find that $Q_{ij}(g_{i},g_{j})$ is the posterior probability that the edge ($i$, $j$) exists and is different depending on whether users $i,j$ belong in the same community($g_{i}=g_{j}=r$) or not ($g_{i}=r, g_{j}=s$, $r \neq s$) with $r,s \in g$:
\small
\begin{equation}
{Q_{ij}(r,r)=\dfrac{p \alpha^{M_{ij}\sigma_{ij}} (1-\alpha)^{M_{ij}(1-\sigma_{ij})}}{p \alpha^{M_{ij}\sigma_{ij}} (1-\alpha)^{M_{ij}(1-\sigma_{ij})} + (1-p) \beta^{M_{ij}\sigma_{ij}} (1-\beta)^{M_{ij}(1-\sigma_{ij})}}},
\label{eq23sbm}
\end{equation}
\vspace{-0.6cm}
\small
\begin{equation}
{Q_{ij}(r,s)=\dfrac{q \alpha^{M_{ij}\sigma_{ij}} (1-\alpha)^{M_{ij}(1-\sigma_{ij})}}{q \alpha^{M_{ij}\sigma_{ij}} (1-\alpha)^{M_{ij}(1-\sigma_{ij})} + (1-q) \beta^{M_{ij}\sigma_{ij}} (1-\beta)^{M_{ij}(1-\sigma_{ij})}}}.
\label{eq23sbmb}
\end{equation}
\normalsize
Notice that for $M_{ij}=0$, $Q_{ij}(g_{i}, g_{j})$ becomes equal to the prior probability $p$ if $g_{i} = g_{j}$ and equal to $q$ if $g_{i} \neq g_{j}$. Next, to maximize the likelihood in terms of the parameters we find:
 %Moreover, from \eqref{eq9} we observe that $q(\textbf{A}, \textbf{g})$ is the posterior probability distribution over all possible networks \textbf{A} and communities \textbf{g}, namely ${P}(\textbf{A}, \textbf{g}, \theta \text{ }| \text{ } \Tc)$, when $Y_{ij}$ is replaced by its expected value $M_{ij}\sigma_{ij}$.
\begin{align}
\alpha = \dfrac{\sum_{i\neq j} M_{ij} \sigma_{ij}Q_{ij}(g_{i},g_{j})}{\sum_{i\neq j} M_{ij} Q_{ij}(g_{i},g_{j})},
\label{eq19b}
\end{align}
\vspace{-0.3cm}
\begin{align}
\beta = \dfrac{\sum_{i\neq j}M_{ij}\sigma_{ij}(1-Q_{ij}(g_{i},g_{j}))}{\sum_{i\neq j}M_{ij}(1-Q_{ij}(g_{i},g_{j}))}, \label{eq41b}
\end{align}
\vspace{-0.3cm}
\begin{align}
p = \dfrac{1}{\sum_{i\neq j}\mathbf{1} (g_{i} = g_{j})} {\sum_{i\neq j, g_{i} = g_{j}} Q_{ij}(g_{i},g_{j}}),
\label{eq21b}
\end{align}
\vspace{-0.3cm}
\begin{align}
q = \dfrac{1}{\sum_{i\neq j}\mathbf{1} (g_{i} \neq g_{j})} {\sum_{i\neq j, g_{i} \neq g_{j}} Q_{ij}(g_{i},g_{j}}).
\label{eq21c}
\end{align}
To find the diffusion probabilities $\sigma_{ij}$ we must solve the following linear optimization problem:
\begin{equation}
\max_{\boldsymbol{\sigma}} \sum_{i \neq j}\sigma_{ij}(W_{ij} - \lambda c) \\
\label{eq17asbm}
\end{equation}
\vspace{-0.3cm}
\begin{equation}
\text{s.t. } \boldsymbol{\sigma} \in {F}_{\boldsymbol{\sigma}}, \nonumber \\
\end{equation}
\vspace{-0.3cm}
\small
\begin{equation}
\small \text{where } \small W_{ij} = M_{ij}\left(Q_{ij}(g_{i},g_{j})\log\frac{\alpha}{1-\alpha} + (1-Q_{ij}(g_{i},g_{j}))\log\frac{\beta}{1-\beta}\right), \nonumber \\
\end{equation}
\vspace{-0.4cm}
\begin{equation}
\lambda > 0 \text{ some given penalty for regularisation, and $c = \max_{(i,j) \in W} W_{ij}$. \nonumber}
\end{equation}
The final CEM-* algorithm, when we choose the SBM prior is shown in Algorithm \ref{algo1}. It iterates between finding an optimal value for $q$, via the $Q_{ij}$ values, and then holding it constant to maximize the likelihood (the right-hand side of \eqref{eq7f} in Appendix B) with respect to $\theta=$\{$\alpha, \beta, p, q, \boldsymbol{\sigma}$\} (M-step). We underline that the updates of the $Q_{ij}$ values that are essential for the E-step require the knowledge of the communities participation vector \textbf{g}. It is updated in each iteration as follows: we generate first a graph from the current $Q_{ij}$ estimations, by drawing an edge whenever $Q_{ij} > 0.5$. To get the updated vector \textbf{g}, we apply to the generated graph the Louvain method, which returns a single community label for each user node (Blondel et al. \citeyear{c1}). Our algorithm converges when the L2 norm of improvement $||\textbf{Q}_{new} - \textbf{Q}_{old} || $ falls under some threshold $\epsilon$ that we choose in advance, where \textbf{Q} is the matrix with the $Q_{ij}$ values.
\begin{figure*}[]
	\includegraphics[width=1\linewidth]{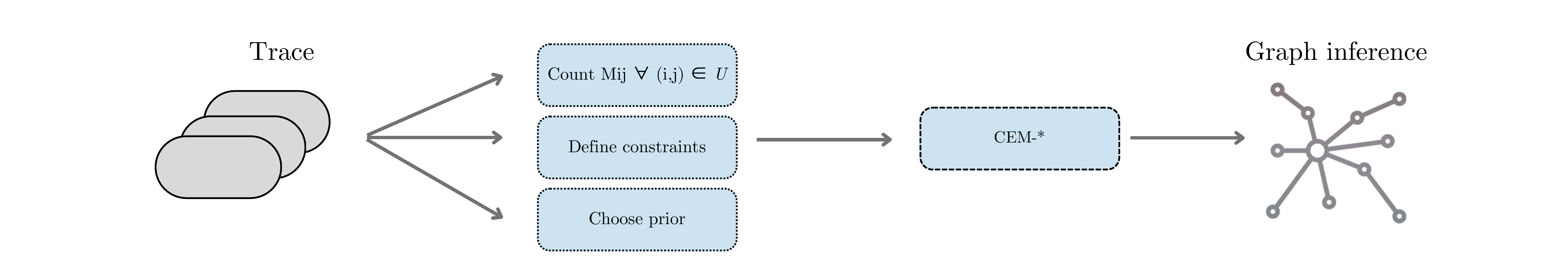}
	\caption{Framework of Constrained-EM}
	\label{framework}
\end{figure*}

\begin{algorithm}
	\caption{CEM-*}
	\begin{algorithmic}[1]
		\renewcommand{\algorithmicrequire}{}
		\renewcommand{\algorithmicensure}{}
		\REQUIRE \textbf{Input:} \$PRIOR, $\Cc$, $\Uc$, $M_{ij}$ $\forall (i,j) \in \Uc$ \\
		\ENSURE \textbf{Output:} $Q, \alpha, \beta, \rho, \boldsymbol{\sigma}$
		\\ t = 0
		\\ \textit{Random Initialisation}: $\alpha_{t}, \beta_{t}, \rho_{t}, \boldsymbol{\sigma}_{t}$
		\IF{\$PRIOR = ER}
		\REPEAT
		\STATE t $\mathrel{+}=$ 1
		\STATE $Q_{t}$ = update $Q(\alpha_{t-1}, \beta_{t-1}, \rho_{t-1}, \boldsymbol{\sigma}_{t-1}$) \text{ using \eqref{eq23ER}}
		\STATE $\alpha_{t}$ = update $\alpha$($Q_{t},\boldsymbol{\sigma}_{t-1}$) using \eqref{ab}
		\STATE $\beta_{t}$ = update $\beta$($Q_{t},\boldsymbol{\sigma}_{t-1}$) using \eqref{ab}
		\STATE $\rho_{t}$ = update $\rho$($Q_{t}$) using \eqref{eq21}
		\STATE $\boldsymbol{\sigma}_{t}$ = update $\boldsymbol{\sigma}$($Q_{t}, \alpha_{t-1}, \beta_{t-1}$) using \eqref{eq17a}
		\UNTIL convergence
		\ELSIF{ \$PRIOR = SBM}
		\STATE \textit{Random Initialisation}: \textbf{g$_{t}$} \\
		\REPEAT
		\STATE t $\mathrel{+}=$ 1
		\STATE \tiny $Q_{t}$ = update $Q(\alpha_{t-1}, \beta_{t-1}, p_{t-1}, q_{t-1}, \boldsymbol{\sigma}_{t-1}$,\textbf{g$_{t-1})$} \text{using \eqref{eq23sbm}, \eqref{eq23sbmb}}
		\normalsize
		\STATE $\alpha_{t}$ = update $\alpha$($Q_{t},\boldsymbol{\sigma}_{t-1}$) using \eqref{eq19b}
		\STATE $\beta_{t}$ = update $\beta$($Q_{t},\boldsymbol{\sigma}_{t-1}$) using \eqref{eq41b}
		\STATE $p_{t}$ = update $\rho$($Q_{t}$) using \eqref{eq21b}
		\STATE $q_{t}$ = update $\rho$($Q_{t}$) using \eqref{eq21c}
		\STATE $\boldsymbol{\sigma}_{t}$ = update $\boldsymbol{\sigma}$($Q_{t}, \alpha_{t-1}, \beta_{t-1}$) using \eqref{eq17asbm}
		\STATE \textbf{g$_{t}$} = LOUVAIN($Q_{t}$)
		\UNTIL convergence
		\ENDIF
	\end{algorithmic}
	\label{algo1}
\end{algorithm}
\begin{table*}[htbp]
	\centering
	\caption{Trace and ground truth statistics for the synthetic and real-world data.}
	\resizebox{0.46\linewidth}{!}{
		\begin{tabular}{lll}
			\hline
			\textbf{Trace statistics} & \textbf{\small Synthetic} & \textbf{\small \#Élysée2017fr} \\
			\hline
			\addlinespace[0.1cm]
			{Time-span} & 17,459 time-steps & 6 months\\
			{\# tweets} & 1,709 & 293,405\\
			{\# retweets} & 24,347& 1,605,059 \\
			{\# users} & 100& 11,521\\
			{\% users with \# tweets > 0} & 27.00 & 70.74 \\
			{\% users with \# retweets > 0 } & 87.00 & 96.45\\
			{\% user pairs with $M_{ij}>0$} & 78.10 & 5.21 \\
			\hline
	\end{tabular}}
	\quad
	\hspace{0.8cm}
	\resizebox{0.46\linewidth}{!}{
		\begin{tabular}{lll}
			\hline
			\textbf{Ground-truth graph} & \textbf{\small Synthetic} & \textbf{\small \#Élysée2017fr} \\ \hline
			\addlinespace[0.1cm]
			{\# edges} & 158& 1,555,718\\
			{\% intra-edges(labeled)} & 63.92 (101) & 84.29 (1,311,463)\\
			{\% inter-edges(labeled)} & 36.08 (57) & 15.71 (244,255) \\
			\addlinespace[0.1cm]
			\hline
			\addlinespace[0.1cm]
			{\% edges with $M_{ij}>0$} & 99.36 (155) & 45.23 (703,682) \\
			{\% intra-edges with $M_{ij}>0$} & 100.00 (101) & 50.13 (657,389) \\
			{\% inter-edges with $M_{ij}>0$} & 98.25 (56) & 18.95 (46,293) \\
			\addlinespace[0.1cm]
			\hline
	\end{tabular}}
	\label{tab:statistics}
\end{table*}

\section{Methodology}
\subsection{Datasets}
The general framework of CEM-* for both priors is shown in Fig. \ref{framework}. To evaluate our two methods CEM-er and CEM-sbm against the ground truth and compare them with existing methods, we will use two different datasets: a synthetic and a real-world one. The synthetic dataset that we create aims to illustrate our method's efficiency when the trace includes sufficient information about the interactions between users. As we will show later, this is not always the case with real-world traces coming from OSNs, which can make the inference task even more challenging.

\subsubsection{Synthetic dataset}
For the generation of synthetic social media data, we follow the code found in Giovanidis et al. (\citeyear{bGiovanidis}). We first create a set of 100 users each of which has two random activity (posting and reposting) rates. Then, we create an SBM graph between the users, with 7 different partitions of varying sizes, that represents the friendship graph of the network. Users in the same group are connected with probability $p=0.06$ and users of different groups are connected with probability $q=0.007$. Each subgraph corresponding to a group is a random Erdős–Rényi with connection probability $p$. For each user, we generate a set of random timestamps, that increase according to an exponential distribution that depends on their activity rates. These timestamps represent the times they posted or reposted something. We generate a set of 100,000 timestamp-user-activity instances in total that we call the {Events} set.

Additionally, we assume that each user has a \text{Newsfeed} that can hold up to 10 posts and reposts from their followees. Based on the friendship graph and the Events set, we simulate a set of interactions between the users according to the following scheme: when a user $i$ visits their Newsfeed, they repost randomly one of the 10 entries made by their followees. A new entry on the Newsfeed list will push out an older entry of a random position. The Newsfeeds of the users that follow user $i$ will then be updated accordingly. Of course, in reality, users on a social media platform may show a preference towards a specific account or topic, or even repost something outside of the scope of their followees. The random uniform selection, however, makes the simulation collect sufficient information for all the edges in the friendship graph.

The simulation generates a social media trace from which we can extract all the quantities that are necessary for our method, as presented in Section \ref{preprocess}. The detailed statistics of the synthetic dataset can be found in Table \ref{tab:statistics}. The table on the left shows the statistics of the trace, whereas the table on the right shows the statistics of the ground truth graph. This is the graph that we will be trying to infer. The intra-edges refer to the edges inside a community, whereas inter-edges refer to the edges between different communities.
\subsubsection{Real-world data: the \#Élysée2017fr dataset}
\label{rldata}
\begin{figure}[!]
	\label{figantoine5}
	\vspace{-0.1cm}
	\begin{subfigure}{\columnwidth}
		\centering
		\begin{subfigure}{0.55\columnwidth}
			\centering
			\includegraphics[width=0.9\columnwidth]{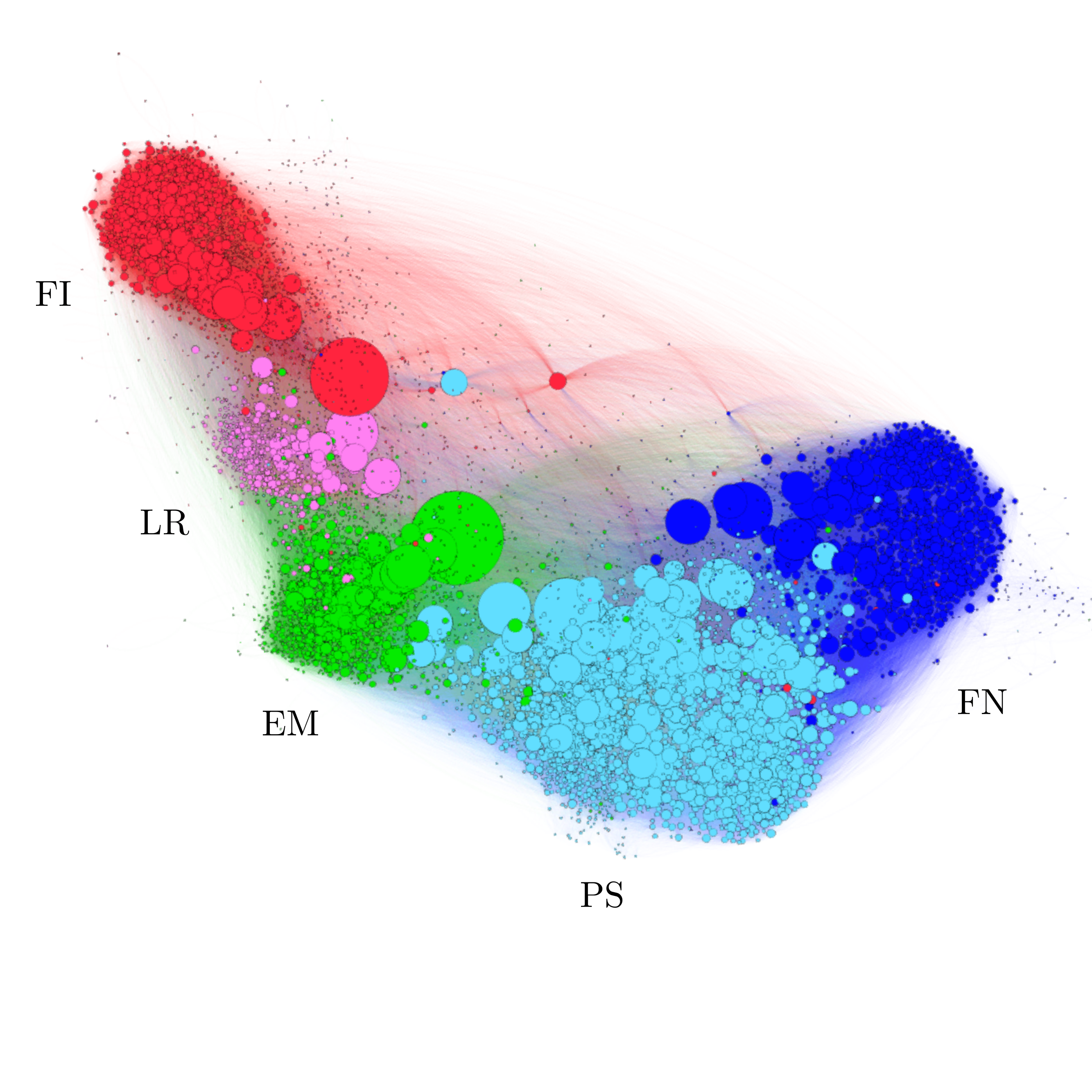}
			\vspace{-0.5cm}
			\caption{Friendship network}
			\label{figgephi}
		\end{subfigure}%
		\hspace{-1cm}
		\begin{subfigure}{.55\columnwidth}
			\centering
			\centering
			\includegraphics[width=0.9\columnwidth]{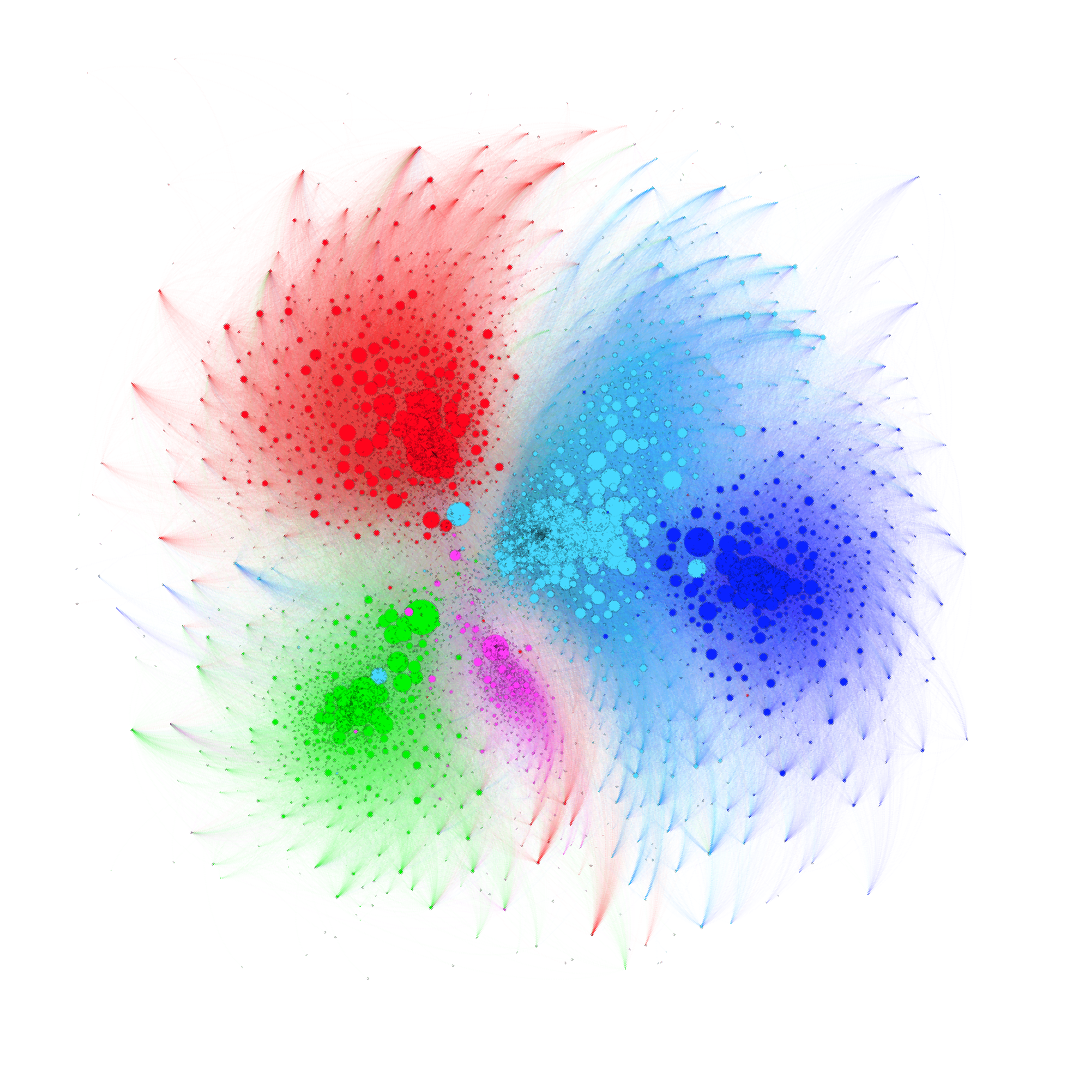}
			\vspace{-0.5cm}
			\caption{Retweet network}
			\label{figgephib}
		\end{subfigure}%
	\end{subfigure}
	\caption{Networks of the 5 political parties provided by the \#Élysée2017fr dataset.}
\end{figure}
For the evaluation of our method on real-world data, we choose \#Élysée2017fr, a publicly available dataset related to the 2017 French presidential campaign on Twitter (Fraisier et al. \citeyear{bElysee}). It features 2,414,584 tweets and 7,763,931 retweets from 22,853 Twitter profiles discussing the election. Users have been manually annotated by experts with political affiliations expressing support for one of the 5 main competing parties in France:
	\vspace{-0.1cm}
\begin{itemize}
	\item FI: France Insoumise, far-left (Jean-Luc Mélenchon)

	\item PS: Parti Socialiste, left-wing (Benoît Hamon)

	\item EM: En Marche, center (Emmanuel Macron)

	\item LR: Les Républicains, right-wing (François Fillon)

	\item FN: Front National, far-right (Marine Le Pen)
\end{itemize}
The exact timestamps of interactions between users had not been published by the creators of the dataset, therefore we had to crawl them using the Twitter API.
On top of that, we collected the follower-followee connections, i.e., the friendship graph of the observed user ids, which had not been provided by the authors. A visual representation of the friendship network that we scraped along with the community participation of each node is shown in Fig. \ref{figgephi}. The network of retweets is shown in Fig. \ref{figgephib}. From these figures, we can see that even though users follow people from other communities (e.g., there are many friendships between the two extreme groups FI and FN), they mostly retweet posts from authors that belong inside their community and they do not interact much with users outside. 

From this trace, we keep only the tweets that have been retweeted by at least one user. Additionally, we remove retweets for which we do not know the author and retweets that have been made more than once by the same user. The statistics of the trace after the above pre-processing along with the statistics of the ground truth graph are shown in Table \ref{tab:statistics}.

\textbf{Insufficient information in a real-world trace.} From Table \ref{tab:statistics} and Fig. \ref{figgephi} and \ref{figgephib}, we notice the main challenge in working with this dataset against the synthetic one: out of the 1,555,718 edges in the underlying friendship graph, only 45.23\% of them have a non-zero $M_{ij}$ value. On the other hand, the synthetic trace includes information for more than 99\% of the 158 existing edges. This can be partly because, in reality, users may repost their followees with some preference, instead of randomly selecting posts from their Newsfeed as is the case in the synthetic dataset. Therefore, many users may not appear to interact with retweets even if there is a connection between them in reality. However, given that the absolute numbers of the real-world trace are quite high, we believe that there is sufficient information to work with.
%Given the above, we expect the optimization to produce more errors and take longer to converge in comparison to the synthetic dataset case.

\subsection{Comparison}
\subsubsection{Compared models}
\label{compared}
We compare the graphs inferred by our two models, CEM-er and CEM-sbm, with those generated by the following baseline and state-of-the-art methods:
\begin{itemize}
	
	\item {\textbf{Star}}: a heuristic graph inference method that draws a directed edge from the author of every tweet $s$ in the trace to every user that appears in the corresponding episode $\Ec_{s}$ after them. The graph inferred by Star implies that all the users that have retweeted a tweet are following its author.
	
	\item {\textbf{Chain}}: another heuristic method that generates a single long path between the users in each episode $\Ec_{s}$, according to the timestamps of their interactions with tweet $s$: each path first connects the author of $s$ to the user $i$ that retweeted it first in time. Then, it connects $i$ to the user $j$ who retweeted it second in time, $j$ to the user who retweeted it third, and so on.
	
	\item {\textbf{Saito et al.} (\citeyear{b5})}: a baseline EM-based algorithm that infers the influence probabilities $k_{ij}$ by assuming an Independent Cascade model of diffusion between the users. For comparison, we produce the final graph by drawing an edge $(i,j)$ whenever $k_{ij} > 0.5$.
	
	\item {\textbf{Netinf}} (\citeyear{b2d}): in a similar way to our work, Gomez-Rodriguez et al. identify the graph that most accurately explains the observed infection times of nodes. However, their formulation of the problem is combinatorial and thus NP-hard to solve exactly. Therefore, they suggest finding near-optimal networks using approximation algorithms, by exploiting the submodularity properties of the objective, which, as we will show in the next sections, introduces computation-time and precision issues. In contrast, we devise a continuous linear expression based on the trace, which allows us to find efficiently the exact solution to an LP optimization problem.
	
	As explained by the authors, when the activity rates are not the same for all users, the performance of the model worsens. Therefore, we expect Netinf to perform worse than CEM-* in more realistic settings such as these of the synthetic dataset, in which users have different activity rates. It should be noted that Netinf requires that we set in advance the parameter $k$, which is the number of edges that we want to infer. For comparison, we set $k$ equal to the number of edges of the corresponding ground truth graph.
	
	\item {\textbf{Newman} (\citeyear{b7})}: a more recent EM-based algorithm  that we introduced in Section \ref{intro}. As mentioned before, our algorithm is an extension of the EM formulation provided in Newman's work. The algorithm is not designed to consider hidden paths between users, thus it is not guaranteed that the inferred networks will be feasible. For evaluation, we derive a graph by drawing an edge $(i,j)$ whenever the friendship probability $Q_{ij}$ for a user pair $(i,j)$ estimated by this method is greater than $0.5$.
	
	\item {\textbf{Peixoto}} (\citeyear{b12}): a state-of-the-art non-parametric Bayesian method that infers posterior distributions from trace observations using a stochastic block model as a prior. As is the case with our CEM-sbm model, it performs community detection together with network reconstruction. Unlike us, however, during the inference process, the model performs sampling using a Markov Chain Monte Carlo procedure and accepts a solution with a Metropolis-Hastings probability. As demonstrated next, this negatively impacts the computation time of the optimization.
\end{itemize}
\subsubsection{Comparison metrics}
\label{metrics}

The directed edges inferred by each inference method translate to the existence of follower-followee relationships between the respective user nodes. To evaluate and compare them against the ground truth, we will look at the following aspects:
\begin{enumerate}
	
	\item \textbf{Results of CEM-* given different trace sizes and values of hyperparameter $\boldsymbol \lambda$.} Firstly, we check how different trace sizes change the corresponding results of our method. For example, by choosing only the first 10,000 lines of the synthetic trace, we obtain information for around 65\% out of the $N \dot (N-1)=9,900$ possible user pairs, whereas the whole trace ($=$ 100,000 lines) informs us on about 78\% of the pairs. We see therefore that as we choose more trace lines from the input, we get more information between users in terms of tweets and retweets (with diminishing returns). In general, we expect the performance of our model to improve with the increasing size of the trace.
	
	\item \textbf{Feasibility of the trace.} We evaluate each method presented in Section \ref{compared} in terms of feasibility. Given the ground truth graph, we check how many episodes are feasible, according to our definition of feasibility provided in Section \ref{feas}. 
	
	\item \textbf{Prediction performance.} When the ground truth is available, we can treat the output of the inference as a binary classification task between existing and non-existing edges. We, therefore, choose Precision, Recall, and AUC scores as metrics for evaluation and comparison. These metrics are used frequently to measure prediction success in similar classification tasks. Precision refers to the percentage of true positive friendships inferred out of all the predicted ones, and Recall quantifies the percentage of true positive friendship edges inferred out of all the edges that are positive in the ground truth. The AUC score is the area under the ROC curve that represents the tradeoff between Recall (true positive rate) and Specificity (false positive rate), not to be mixed with the true and false positive utilization rates $\alpha$ and $\beta$ in the parameters set $\theta$ of CEM-*. It is a measure of separability and quantifies how well the model can distinguish between classes. 
	
	%Depending on the application and the desired outcome, one can choose which of these metrics is more indicative of a successful inference output.
	
	\item \textbf{Inferred network metrics.} Additionally, we look into different network measures of the inferred graph (e.g., average degree, diameter, connected components, etc), and compare them to these of the ground truth graph. These measures can be indicative of how much the inferred graph resembles the properties of a general real-world graph (in cases when the ground truth is not available).
	
	\item \textbf{Detection of communities.} A useful by-product of our CEM-sbm network reconstruction method is the community detection task. Therefore, we check to what extent the inferred communities resemble the real ones presented in the ground truth. Since a node can only belong to one community, we wish to verify whether the different pairs of users belonging to the same or different communities are the same in the ground truth. The method for the evaluation and comparison is the following: we first generate a graph for each model as described in Section \ref{compared} and then apply on it the Louvain method for community detection (Blondel et al. \citeyear{c1}). The detected clusters are then used to calculate the F1-score as follows: we look at each possible user pair and if the users belong to the same community we label the edge with 1 (positive class), otherwise with 0 (negative class). We do the same for the ground truth (with the Louvain labels). From the true/false positive, and true/false negative rates we measure the F1-score, which combines Precision and Recall. In addition, we estimate the values of $p$ and $q$ between the communities in the inferred graph and compare them to the real ones.
\end{enumerate}

\begin{figure*}[]
	\begin{subfigure}{\columnwidth}
		\begin{subfigure}{.50\columnwidth}
			\includegraphics[width=1\columnwidth]{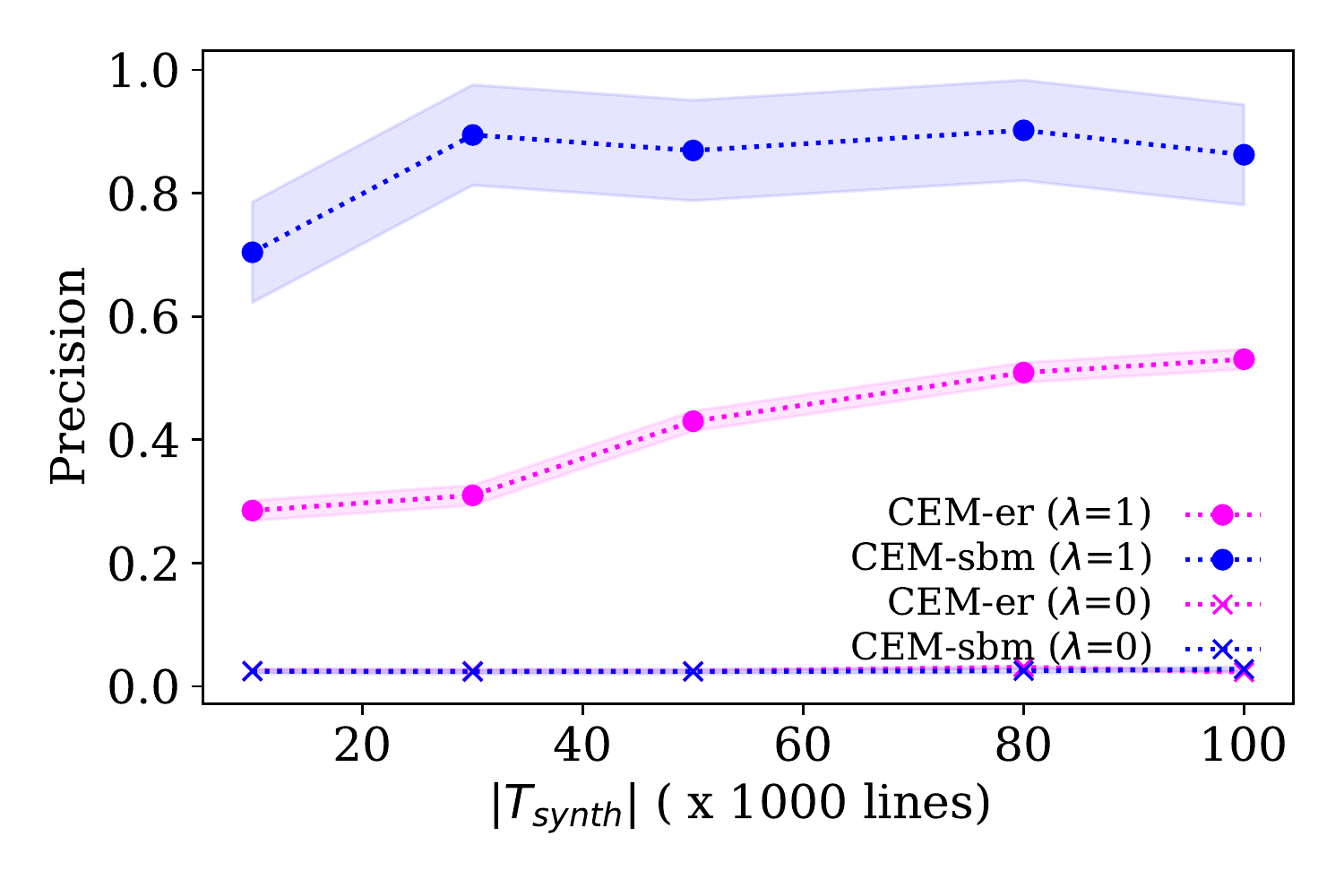}
		\end{subfigure}%
		\hfill
		\begin{subfigure}{.50\columnwidth}
			\centering
			\includegraphics[width=1\columnwidth]{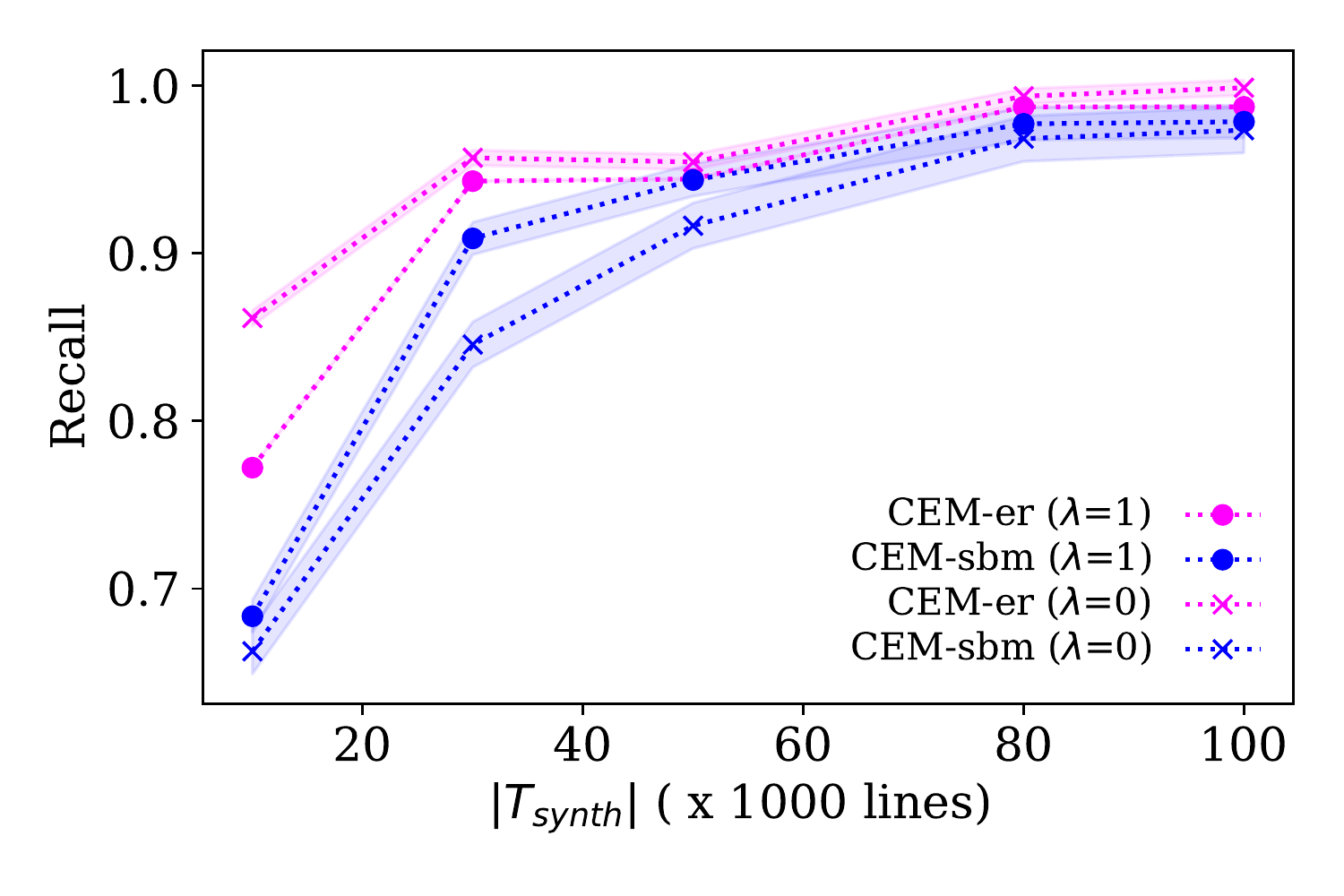}
			
		\end{subfigure}%
		\caption{Results for different sizes of the synthetic trace $T_{synth}$.}
		\label{fig:synth-eval1}
	\end{subfigure}
	\hfill
	\begin{subfigure}{\columnwidth}
		\begin{subfigure}{.50\columnwidth}
			\centering
			\includegraphics[width=1\columnwidth]{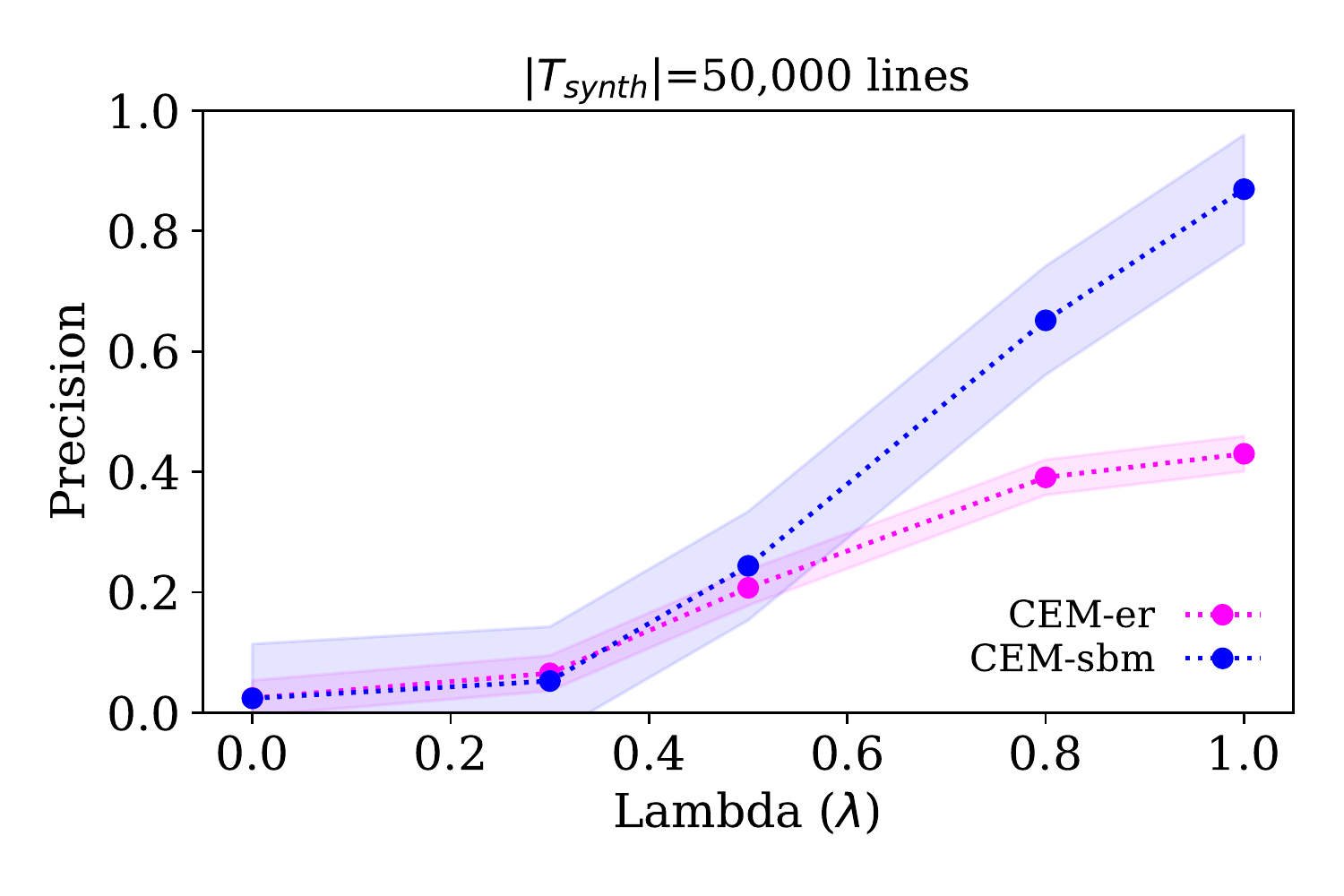}
			
		\end{subfigure}%
		\hfill
		\begin{subfigure}{.50\columnwidth}
			\centering
			\includegraphics[width=1\columnwidth]{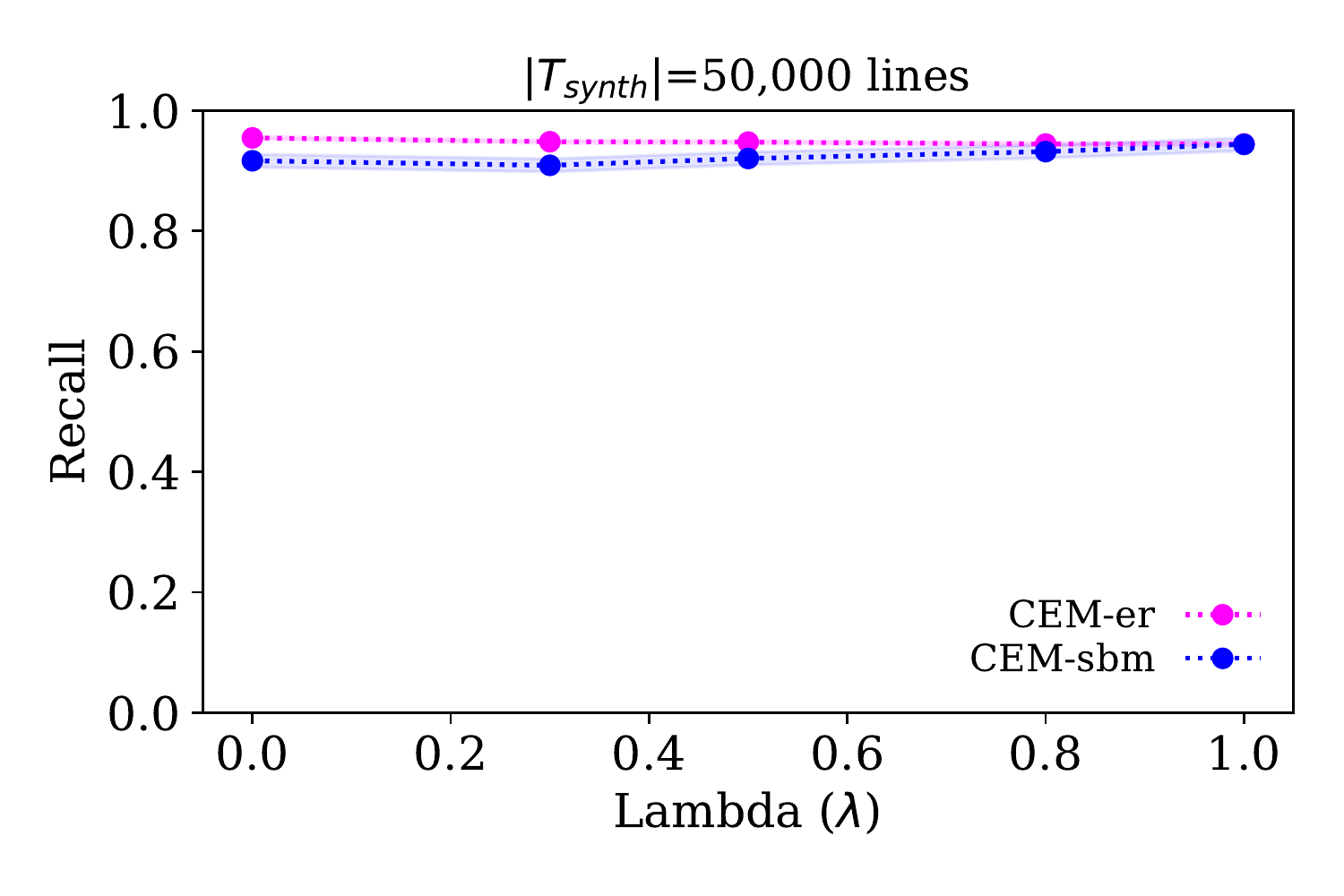}
			
		\end{subfigure}%
		\caption{Results for different values of $\lambda \in (0,1)$ given |$T_{synth}$| = 50,000 lines.}
		\label{fig:synth-eval2}
	\end{subfigure}%
	\caption{Precision given Recall of CEM-er and CEM-sbm applied on the synthetic dataset.}
\end{figure*}

\subsection{Experimental settings}
We run the experiments on a virtual machine with 40 vCPUs and 256 GB RAM. For the solution to the optimization problem, we configure a Gurobi solver through PuLP\footnote{https://pypi.org/project/PuLP/}, an open-source linear programming library for Python, using the dual simplex optimization method. The parameters set $\theta_{1}=\{\alpha, \beta, r, \boldsymbol \sigma \}$ and $\theta_{2}=\{\alpha, \beta, p, q, \boldsymbol \sigma \}$ for CEM-er and CEM-sbm respectively are initialized uniformly at random in the range $[0,1]$. As a convergence criterion for the optimization we choose the L2 norm of the difference between the values of $\textbf{Q}$, i.e., $||\textbf{Q}_{new} - \textbf{Q}_{old} || < \epsilon$, where the threshold $\epsilon$ is set equal to $0.001$. Finally, to generate the unknown friendship graph $G$, we round up all edges with $Q_{ij} > 0.5$ to 1, and the rest are set to $0$. We run the experiments 10 times and report the average results.

\section{Experiments on synthetic data}

\begin{table}[htbp]
	\vspace{-0.3cm}
	\centering
	\caption{Converged parameters for |$T_{synth}$| = 50,000 lines.}
	\resizebox{0.7\columnwidth}{!}{%
		\begin{tabular}{lll}
			\hline
			\addlinespace[0.1cm]
			{\text{CEM-* parameters}} &  {1-$\alpha^*$}  & {$\beta^*$}  \\ \hline
			\addlinespace[0.1cm]		
			{CEM-er ($\lambda=0$)}  & 1-(7e-11) & 1.74e-12 \\
			{CEM-er ($\lambda=1$)}  & 1-(2e-10) & 1.54e-13 \\
			\addlinespace[0.1cm]
			{CEM-sbm ($\lambda=0$)}  & 1-(7e-11) & 1.65e-12   \\
			{CEM-sbm ($\lambda=1$)}  & 1-(9e-11) & 3.95e-15   \\ 
			\hline 
	\end{tabular}}
	\label{tab:par-synth}
	\vspace{-0.3cm}
\end{table}

\begin{figure*}[]
	\centering
	\begin{subfigure}{0.3\columnwidth}
		\includegraphics[width=1.2\linewidth]{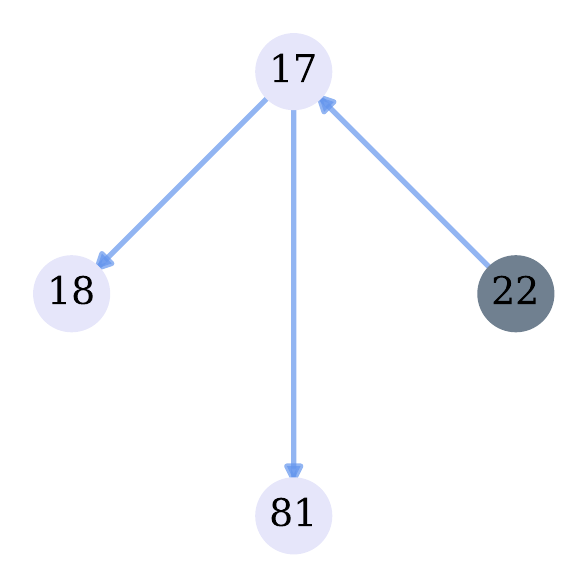}
		\caption{Synthetic (true)}
	\end{subfigure}%
	\hfill
	\begin{subfigure}{0.3\columnwidth}
		\includegraphics[width=1.2\linewidth]{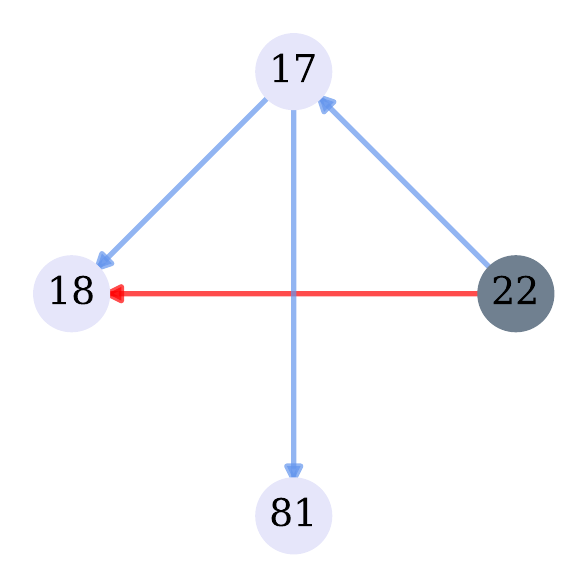}
		\caption{CEM-er($\lambda=1$)}
	\end{subfigure}%
	\hfill
	\begin{subfigure}{0.3\columnwidth}
		\includegraphics[width=1.2\linewidth]{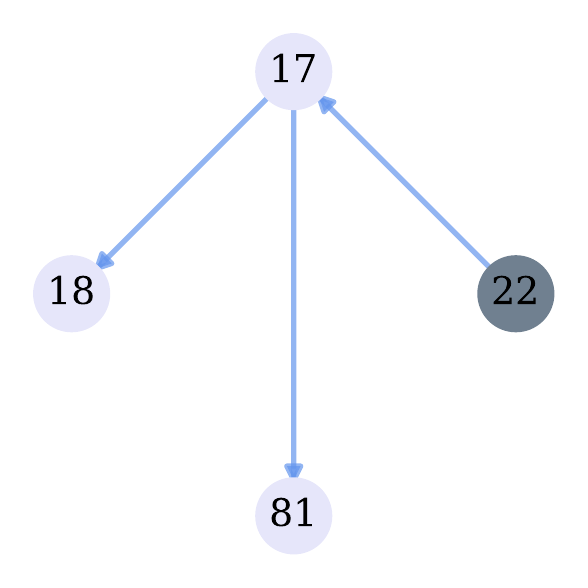}
		\caption{CEM-sbm($\lambda=1$)}
	\end{subfigure}%
	\hfill
	\begin{subfigure}{.3\columnwidth}
		\includegraphics[width=1.2\linewidth]{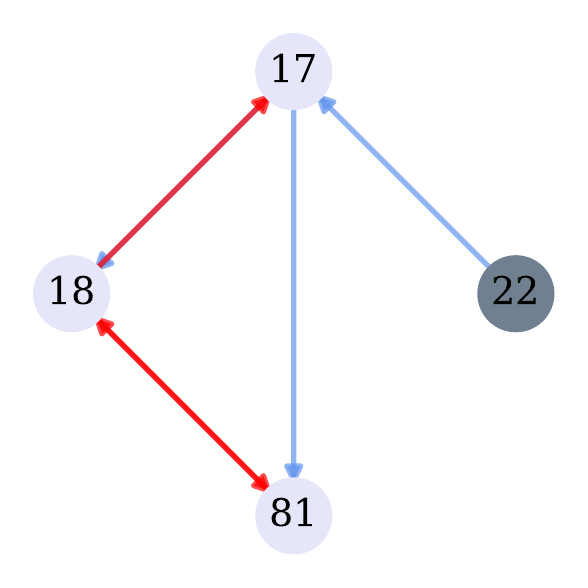}
		\caption{Chain}
	\end{subfigure}%
	\hfill
	\\
	\centering
	\begin{subfigure}{0.3\columnwidth}
		\includegraphics[width=1.2\linewidth]{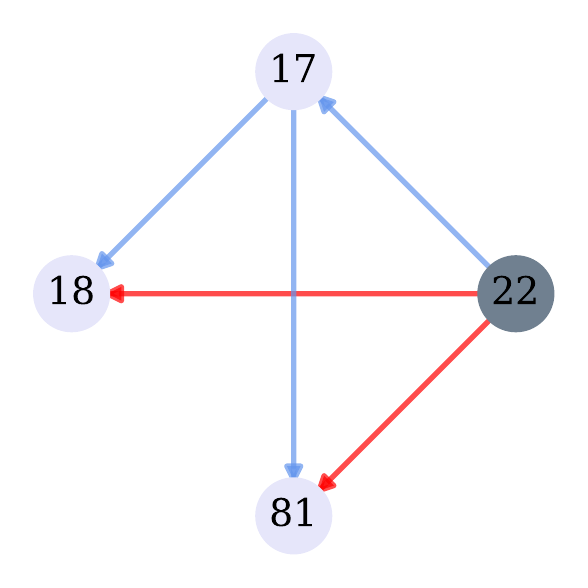}
		\caption{Star}
	\end{subfigure}%
	\hfill
	\begin{subfigure}{0.3\columnwidth}
		\includegraphics[width=1.2\linewidth]{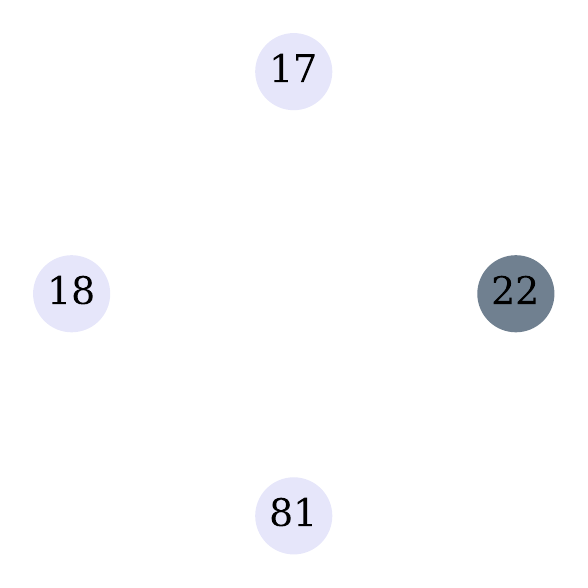}
		\caption{Newman (\citeyear{b7}) / Saito et al. (\citeyear{b5})}
	\end{subfigure}%
	\hfill
	\begin{subfigure}{.3\columnwidth}
		\includegraphics[width=1.2\linewidth]{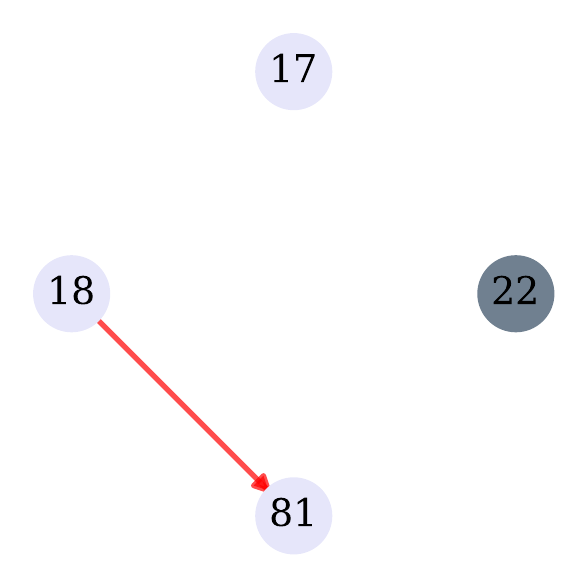}
		\caption{Netinf (\citeyear{b2d})}
	\end{subfigure}%
	\hfill
	\begin{subfigure}{.3\columnwidth}
		\includegraphics[width=1.2\linewidth]{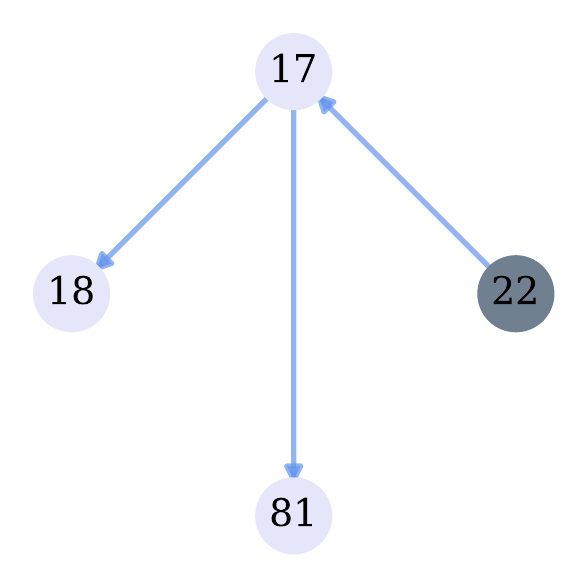}
		\caption{ Peixoto (\citeyear{b12})}
	\end{subfigure}%
	\caption{Comparison of the propagation graph inferred by each method for an episode $\Ec_{s}=\{22, 17, 18, 81\}$ from the synthetic trace when |$T_{synth}$| = 50,000 lines. Each graph shows the real propagation of the tweet $s$ from its author (user 22) to every other user that retweeted it. Blue arrows stand for true positive edges and red arrows stand for false positive ones.}
	\label{fig:example-synthetic}
\end{figure*}

\subsection{Results of our method (CEM)}
\textbf{Values of parameters.} The converged parameters of both our methods, CEM-er and CEM-sbm are shown in Table \ref{tab:par-synth}. In the first column, we show $(1-\alpha^*)$ to be precise about how small the distance is from the maximum value of $\alpha^*$ that is equal to 1. We observe that in every case $\alpha$ is close to 1. This means that there is an almost $100\%$ probability that a post propagated through an edge present in the network we inferred. On the other hand, the small values of $\beta$ suggest that the number of false positive utilized edges is close to zero. This suggests that a post from the trace always propagates through an edge that has been inferred.

\textbf{Different sizes of input.} Figure \ref{fig:synth-eval1} shows the relation of Precision and Recall given trace sizes that range from 10,000 to 100,000 lines. As we observe, the larger the trace, the higher the value of Recall. This was expected since bigger traces give more information which helps us derive more underlying edges. Precision presents relatively stable behavior and is higher ($=$0.869) when $\lambda=1$. Overall, we see that CEM-sbm has higher performance than CEM-er in terms of Precision which reaches up to 0.869 when $\lambda=1$, and a slightly worse, but still competitive performance in terms of Recall (reaching up to 0.944 for $\lambda=1$ whereas CEM-er can reach up to $0.954$ for $\lambda=0$). We conclude therefore that CEM-sbm is much more precise than CEM-er in the case of the synthetic dataset and can also retrieve most of the underlying edges.

\textbf{Different values of the hyperparameter $\boldsymbol \lambda$}. In Fig. \ref{fig:synth-eval2} we can see more clearly how the choice of the hyperparameter $\lambda$ inside the optimization objective (Eq. \ref{eq17a} and Eq. \ref{eq17asbm}) affects the precision of inference: for $\lambda=0$ we get very low Precision ($=0.024$) regardless of the prior since we infer the largest number of edges possible according to the objective, which in turn results to more false positive edges. However, in this case, the Recall value is at its highest (for example $0.954$ in the case of CEM-er). In contrast, for $\lambda=1$, we infer a graph with the smallest number of edges possible given the constraints and thus we get a considerably better Precision ($=0.869$, CEM-sbm). The Recall value, in this case, is still high ($=0.944$). This can be linked to the rich information that is provided in the synthetic trace but can also be indicative of the good prediction probabilities of our method: we manage, with the help of the constraints, to infer the smallest set of edges possible (by setting $\lambda=1$), that is precise and at the same time retrieves almost the entire ground truth graph.

\textbf{Difference between priors.} Choosing $\lambda=1$ we can tell the difference between the ER and SBM priors - the latter is more efficient in the task of inferring more true positive and less false positive connections between the users, achieving a Precision close to 0.9. This suggests that, in CEM-sbm, the use of the priors $p$ and $q$ in Eq. \ref{eq21b} and Eq. \ref{eq21c} depending on whether an $(i,j)$ user pair belongs in the same community or not, instead of the use of a global parameter $r$ (as in Eq. \ref{eq21}) that is unaware of any community structure, can greatly improve the prediction performance of the optimization when there are communities in the real graph. Additionally, as shown in Table \ref{tab:communities-synth} and as we will show later in more detail, CEM-sbm can detect the underlying communities much better than CEM-er: the estimated $p,q$ values of the graph derived by CEM-sbm for $\lambda=1$ are much closer to reality ($p=0.063$ and $q=0.006$), with small relative errors ($\epsilon_{p}=0.05$ and $\epsilon_{q}=0.143$), while the F1-score is almost optimal ($=0.961$). This is a substantial improvement over the F1-score provided by CEM-er ($=0.419$).
\begin{table*}[htpb]
	\centering
	\caption{Performance of different methods on a synthetic dataset with |$T_{synth}$| = 50,000 lines as input.}
	\resizebox{0.5\linewidth}{!}{%
		\begin{tabular}{lllll}
			\hline
			\addlinespace[0.1cm]
			{Performance} & {Precision} & {Recall} & {AUC} & {runtime (secs)} \\ \hline
			\addlinespace[0.1cm]
			{Star} & 0.141 & \textbf{0.956} & {0.931} & \textbf{1.0}\\
			{Chain} & 0.033 & \underline{0.955} & 0.752 & \textbf{1.0}\\
			{Saito et al. (\citeyear{b5})} & \textbf{1.0} & 0.051 & 0.525 & \text{3.0}\\
			{Netinf (\citeyear{b2d})} & 0.159 & 0.165 & 0.575 & 2,199.0 \\
			{Newman (\citeyear{b7})} & 0.522 & 0.450 & 0.724 & \text{2.0} \\
			{Peixoto (\citeyear{b12})} & \text{0.643} & 0.924 & \text{0.958} & 3,481.0 \\
			\addlinespace[0.1cm]
			\hline
			\addlinespace[0.1cm]
			{CEM-er ($\lambda=0$)} & 0.024 & \text{0.954} & 0.668 & 8.0 \\
			{CEM-er ($\lambda=1$)} & 0.430 & \text{0.944} & \underline{0.962} & 9.0 \\
			\addlinespace[0.1cm]
			{CEM-sbm ($\lambda=0$)} & 0.024 & 0.916 & 0.650 & \underline{1.4} \\
			{CEM-sbm ($\lambda=1$)} & \underline{0.869} & \text{0.944} & \textbf{0.970} & \text{1.5} \\
			\addlinespace[0.1cm]
			\hline
	\end{tabular}}
	\label{comp-synthetic}
\end{table*}
\begin{table*}[!]
	\centering
	\caption{Network statistics of the graph inferred by each method compared to the ground truth for |$T_{synth}$| = 50,000.}
	%For each graph it shows the number of edges, the average out-degree, the maximum out and in-degrees, the diameter, the average shortest path, and the maximum strongly connected component in terms of percentage of users.
	\resizebox{\linewidth}{!}{%
		\begin{tabular}{lllllllllll}
			\hline
			\addlinespace[0.1cm]
			{Inferred network metrics} & feasibility(\%) &
			\#edges &
			avg out-degree &
			max out-degree &
			max in-degree &
			diameter &
			avg shortest path &
			max scc (\%users)
			\\ \hline
			\addlinespace[0.1cm]
			{Synthetic graph} & 100.00 & 164 & 1.64 & 39 & 15 & 5 & 2.57 & 11 \\
			\hline
			\addlinespace[0.1cm]
			
			Star & \textbf{100.00} & 1,072 & 10.72 & 78 & 24 & 3 & 1.35 & \underline{10} \\
			Chain & \textbf{100.00} & 4,545 & 46.86 & 75 & 71 & 3 & 1.48 & 87 \\
			Saito et al. (\citeyear{b5}) & 2.33 & 8 & 0.50 & 1 & 1 & 1 & 1 & 0 \\
			Netinf (\citeyear{b2d}) & 34.80 & 164$^{*}$ & 2.49 & 9 & \underline{12} & 12 & 4.76 & 24 \\
			Newman (\citeyear{b7}) & 72.29 & \underline{138} & \textbf{1.55} & 77 & 3 & 1 & 1 & 0 \\

			Peixoto (\citeyear{b12}) & \underline{98.02} & 227 & 2.34 & \underline{36} & 11 & 10 & 3.42 & 19 & \\
			\addlinespace[0.1cm]
			\hline
			\addlinespace[0.1cm]
			CEM-er ($\lambda=0$) & \textbf{100.00} & 6,175 $\pm$ 89 & 63.66 $\pm$ 0.92 & 94.5 $\pm$ 0.17 & 96 & 2 & 1.34 & 97

			\\
			
			CEM-er ($\lambda=1$) & \textbf{100.00} & 349 $\pm$ 8 & 3.59 $\pm$ 0.09 & 45.9 $\pm$ 2.03 & \textbf{15.2 $\pm$ 0.2} & \textbf{5} & \underline{2.23} & \underline{10} 	\\
			
			CEM-sbm ($\lambda=0$) & \textbf{100.00} & 6,141 $\pm$ 244& 63.31 $\pm$ 2.52 & 91.5 $\pm$ 2.95 & 92.6 $\pm$ 3.4 & 2.1 & 1.34 & 97

			\\
			
			CEM-sbm ($\lambda=1$) & \textbf{100.00} & \textbf{177 $\pm$ 11} & \underline{1.83 $\pm$ 0.12} & \textbf{41.3 $\pm$ 1.65} & 11.3 $\pm$ 0.15 & \underline{5.2 $\pm$ 0.2} & \textbf{2.61} & \textbf{11.9 $\pm$ 0.6}

			\\ \hline
			
			*\small chosen a priori
			\\
	\end{tabular}}
	\label{tab:stats-synth}
\end{table*}
\begin{figure}
	\centering
	
	\includegraphics[width=0.8\columnwidth]%
	{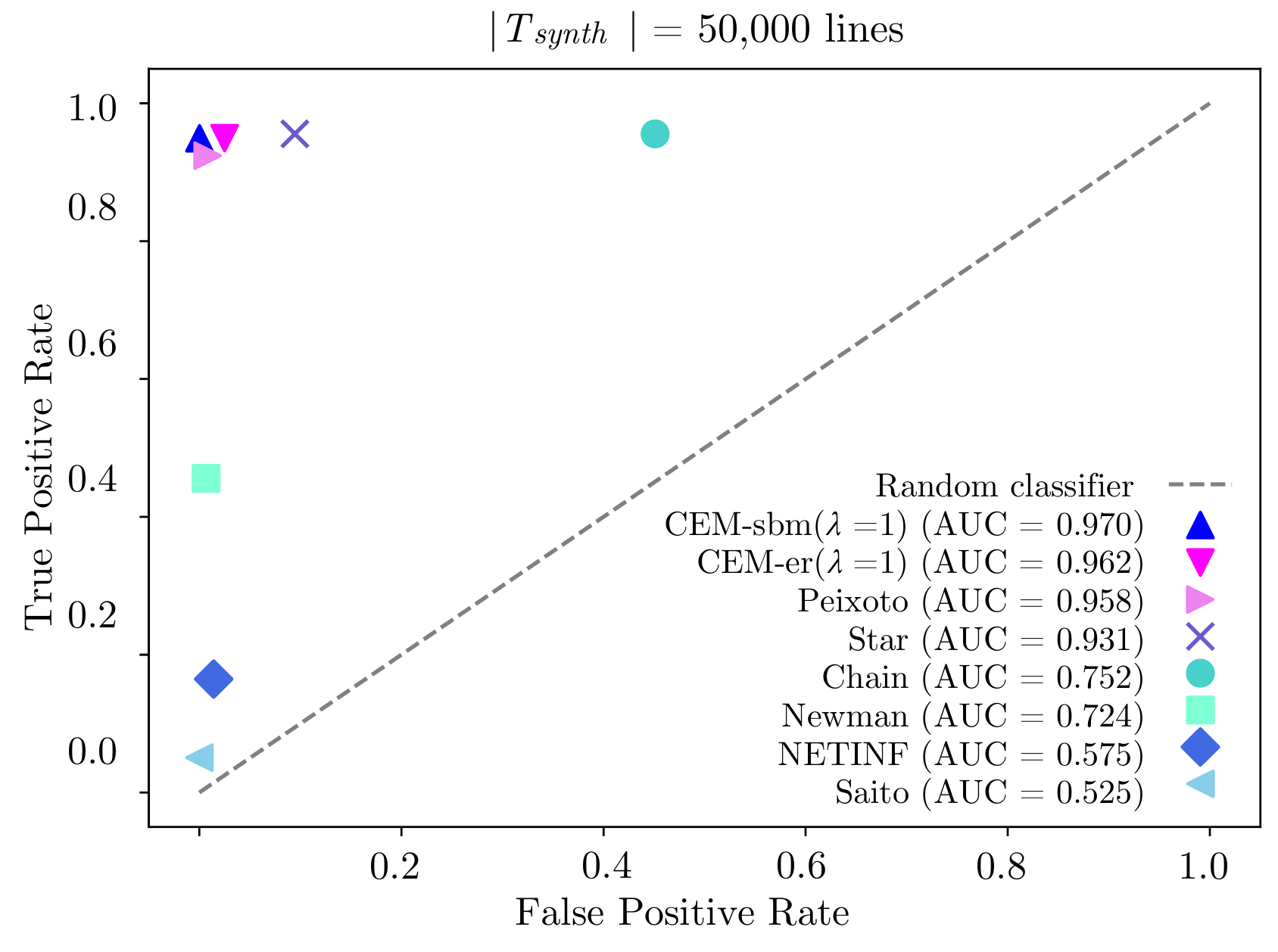}
	\caption{True Positive related to the False Positive rates of each inference model when applied to the synthetic dataset.}
	\label{fig:roc-synth}
	\vspace{-0.4cm}
\end{figure}
\begin{table*}[]
	
	\caption{Performance of community detection for the synthetic graph with |$T_{synth}|$ = 50,000 lines.}
	\hspace{1.6cm}
	\resizebox{0.5\columnwidth}{!}{

		\begin{tabular}{ll}
			\hline
			\addlinespace[0.1cm]
			\text{Label prediction} & {\text{F1-score}} \\ \hline
			\addlinespace[0.1cm]
			
			{Star} & 0.350\\
			{Chain} & 0.421 \\
			{Saito et al. (\citeyear{b5})} & -- \\
			{Netinf (\citeyear{b2d})} & 0.251 \\
			{Newman (\citeyear{b7})} & 0.526 \\
			{Peixoto (\citeyear{b12})} & \underline{0.731} \\
			\addlinespace[0.1cm]
			\hline
			\addlinespace[0.1cm]
			{CEM-er ($\lambda=1$)} &	0.419	\\
			{CEM-sbm ($\lambda=1$)} & \textbf{0.961} \\
			\hline
			\addlinespace[0.44cm]
	\end{tabular}}
	\hspace{2.8cm}
	\resizebox{0.95\columnwidth}{!}{
		
		\begin{tabular}{lllll}
			\hline
			\addlinespace[0.1cm]
			{\text{Community parameters}} & $p_{G_{synth}}$ & |$\epsilon_{p}$| & $q_{G_{synth}}$ & |$\epsilon_{q}$| \\ \hline
			\addlinespace[0.1cm]
			Synthetic graph & \text{0.060}& -- & \text{0.007} & -- \\
			\hline
			\addlinespace[0.1cm]
			
			{Star} & 0.148 & 1.467 & 0.091 & 12 \\
			{Chain} & 0.682 & 10.366 & 0.351 & 49.142 \\
			{Saito et al. (\citeyear{b5})} & 0.500 & 7.333 & -- & --\\
			{Netinf (\citeyear{b2d})} & 0.285 & 3.750 & 0.002 & 0.714 \\
			{Newman (\citeyear{b7})} & \underline{0.043} & \underline{0.283} & \textbf{0.007} & \textbf{0} \\
			{Peixoto (\citeyear{b12})} & 0.112 & 0.866 & 0.006 & \text{0.143}\\
			\addlinespace[0.1cm]
			\hline
			\addlinespace[0.1cm]
			%{CEM-er ($\lambda=0$)} & 0.757 & 0.545 \\
			{CEM-er($\lambda=1$)} & 0.085 & 0.416 & 0.021 & 2.000\\
			
			%{CEM-sbm ($\lambda=0$)} & 0.839 & 0.600\\
			{CEM-sbm($\lambda=1$)} & \textbf{0.063}& \textbf{0.050}
			& \underline{0.006} & \underline{0.143}\\
			\hline
			\addlinespace[0cm]
	\end{tabular}}

	\label{tab:communities-synth}
	
\end{table*}

\subsection{Comparison between methods}
\subsubsection{Propagation subgraph inferred by each model}
For a first understanding of the inner workings of each method that we compare with, we can zoom into the propagation graph inferred for a random episode $\Ec_{s}=\{22, 17, 18, 81\}$ from the synthetic trace (Fig. \ref{fig:example-synthetic}). Each method receives as an input the first 50,000 lines of the original trace that after preprocessing contains 859 tweets and 12,236 retweets. The ground truth tells us that users 18 and 81 have reposted user 17, who had previously reposted directly the author user 22. As we see in Fig. \ref{fig:example-synthetic}, our method CEM-sbm ($\lambda=1$) and Peixoto (\citeyear{b12}) have inferred the propagation graph of the episode correctly. CEM-er ($\lambda=1$) has inferred one more false positive edge from 22 to 18 whereas Star and Chain have inferred two false positive edges. Netinf (\citeyear{b2d}) has inferred only one false positive edge from 18 to 81 whereas the methods by Newman (\citeyear{b7}) and Saito et al. (\citeyear{b5}) have inferred no edge at all. Of course, this is only one example of a subgraph inferred by each method. We are going to see next the performance and statistics of the entire friendships graphs inferred. 

\subsubsection{Performance comparison}

\textbf{Precision, Recall, AUC, and graph statistics.} Firstly, we are comparing CEM-* with the other methods by looking into the graphs and the performance of each model as described in Section \ref{metrics}. More specifically, we will compare the performance of each method in terms of Precision, Recall, and AUC. The results are shown in Table \ref{comp-synthetic} and are combined with observations from each graph's statistics, found in Table \ref{tab:stats-synth}\footnote{The highest value is marked with boldface and the second highest value is underlined. max scc: maximum strongly connected component.}.

From there we observe that the two heuristics, \textbf{Star and Chain}, give 100\% feasible solutions. However, both methods infer graphs with thousands of edges (1,072 and 4,545 edges respectively) and high average out-degrees (10.72 and 46.86) which is very far from reality: the ground truth features only 164 connections with an average out-degree of 1.64. This may result in high Recall and AUC scores but comes at the cost of a very low Precision rate (0.141 and 0.033 respectively, as seen in Table \ref{comp-synthetic}). Additionally, both methods infer graphs with very small average shortest paths ($ < 1.5 $). In contrast, the ground truth has an average shortest path of 2.57 which is closer to the value that we would expect from a real-world Twitter graph to have. Moreover, Chain infers graphs that are too dense, as seen from its maximum strongly connected component (last column, Table \ref{tab:stats-synth}: it includes 87\% of the users, whereas the actual value is only 11\%). The above suggests that, given the synthetic dataset as input, Star and Chain infer graphs that are feasible but demonstrate properties that are far from these of the actual graph, and also, from these of a real-world graph in general.

The method of \textbf{Saito et al. (\citeyear{b5})} is 100\% precise but produces only 8 edges, a very low number for it to be considered a sufficient solution to our problem. Consequently, it presents a very low feasibility rate: it can only explain 2.33\% of the episodes presented in the trace. As a result, its graph properties are far from those of the real graph. For example, the maximum out and in-degrees of the graph are equal to 1, along with the diameter and the average shortest path. Furthermore, the graph inferred by Saito has no strongly connected component and has a very low average out-degree of 0.5.

For the \textbf{Netinf (\citeyear{b2d})} model, we set in advance $k = 164$ as the number of edges that we want to infer, which is equal to the number of edges of the real graph (however such information will not be available in practice and the authors suggest trying different values of $k$ depending on the desired outcome). As we see, the inferred graph has low feasibility of 34.8\% and performs poorly on Precision ($=0.159$), Recall ($=0.165$), and AUC (=$0.575$). This is accompanied by weak graph statistics: it has a relatively low maximum out-degree ($=9$ whereas the real value is 39), the largest diameter out of all the methods ($=12$), and its maximum strongly connected component is more than two times bigger than the real one (it covers 24\% of the users).

The method by \textbf{Newman (\citeyear{b7})} returns a Precision$=0.522$ and Recall$=0.450$ which are values close to the output of a random classifier. However, it infers a graph with 138 edges and an average degree of 1.55 which is close to the real numbers. Still, the diameter, maximum in-degree, and average shortest path values are really small compared to the ground truth. Additionally, it presents no strongly connected component. All in all, the graph is neither feasible (feasibility $=72.29$\%), nor competitive in terms of any performance or statistical metric, which could be due to the fact that it does not consider the hidden paths that exist between users and thus, loses a lot of information that is (indirectly) available in the trace.

The method by \textbf{Peixoto (\citeyear{b12})} is the most competitive out of all the above methods, with 98\% feasibility, Precision $=0.643$ and Recall$=0.924$. Additionally, the graph presents some properties that are similar to the ground truth. For example, as we see in Table \ref{tab:stats-synth}, the derived graph has a maximum out-degree ($=36$) whose value is the second closest to the real one ($=39$). However, it generates almost 40\% more edges and therefore the diameter and the maximum strongly connected component of the graph is almost two times larger than the true one.

To compare with the above, both our methods, CEM-er and CEM-sbm achieve 100\% feasibility across all $\lambda$ values. In addition, CEM-sbm ($\lambda=1$) achieves the highest performance out of all the methods in terms of Precision, Recall, and AUC (=$0.869, 0.944, 0.970$ respectively). Furthermore, we see that the graph inferred by CEM-sbm for $\lambda=1$ has network properties almost identical to the ground truth, followed by the one inferred by CEM-er ($\lambda=1$).

\textbf{Optimization runtime.} On top of the good prediction and graph statistics results, our algorithm is scalable and achieves running times that are close to the times of the heuristics and far lower than other alternatives (last column, Table \ref{comp-synthetic}). CEM-sbm for example runs in less than 1.5 seconds, which is close to the runtimes of Star and Chain. The methods by Newman (\citeyear{b7}) and Saito et al. (\citeyear{b5}) may have similar runtime, but they lose in accuracy. In contrast, Netinf (\citeyear{b7}) and Peixoto (\citeyear{b12}) need more than half an hour to converge and still, as we saw above, their results are not as competitive. This makes our optimization method powerful not only in terms of the accuracy of the prediction but also in terms of the time that is needed to reach a result.

\textbf{ROC curve points of each method}. The Precision and Recall point shown in Table \ref{comp-synthetic} for all methods are also visually illustrated on a 2-dimensional True Positive vs False Positive Rate scale (Fig. \ref{fig:roc-synth}). The upper left corner points correspond to the ideal classifier with AUC$=1$; close to that point we find CEM-sbm ($\lambda=1$), CEM-er ($\lambda=1$), and Peixoto (\citeyear{b12}). Star is close, while the other methods are further away.

\textbf{Detection of communities.} As shown in Table \ref{tab:communities-synth}, our method CEM-sbm ($\lambda=1$) achieves the highest F1-score ($=0.961$) out of all the methods, followed by the method by Peixoto ($=0.731$). Interestingly, the $p$, $q$ parameters of CEM-sbm ($\lambda=1$) are close to these of the ground truth ($p_{G_{synth}}=0.063$ with relative error $|\epsilon_{p}|=0.05$ and $q_{G_{synth}}=0.006$ with relative error $|\epsilon_{q}|=0.143$). Among the other methods, regarding $p$ and $q$, we see that the method by Newman (\citeyear{b7}) presents the lowest relative errors regarding the real values ($0.283$ and $0$ respectively).

\section{Experiments on the \#Élysée2017fr dataset}

Next, we will work with real-world data that, as seen in Section \ref{rldata}, have different properties from the synthetic dataset, making the inference process more challenging.

\begin{table}[htbp]
	
	\centering
	
	\caption{Converged values for parameters $\alpha, \beta$ given |$T_{elysee}$| = 5,000,000 lines as input.}
	
	\resizebox{0.6\columnwidth}{!}{%
		
		\begin{tabular}{lll}
			
			\hline
			
			\addlinespace[0.1cm]
			
			\text{CEM-* parameters} & \textbf{$(1-\alpha^*)$} & \textbf{$\beta^*$} \\ \hline
			
			\addlinespace[0.1cm]
			
			{CEM-er ($\lambda=0$)} & 0 & 1.19e-10 \\
			
			{CEM-er ($\lambda=1$)} & 1.32e-11 & 2.46e-12 \\
			
			\addlinespace[0.1cm]
			
			{CEM-sbm ($\lambda=0$)} & 5.55e-16 & 1.07e-10

			\\
			
			{CEM-sbm ($\lambda=1$)} & 0.004 & 0.001

			\\

			\hline
			
	\end{tabular}}
	
	\label{tab:param-elysee}
	\vspace{-0.5cm}
\end{table}

\begin{figure*}[!]
	\begin{subfigure}{\columnwidth}
		\begin{subfigure}{.50\columnwidth}
			\includegraphics[width=1\columnwidth]{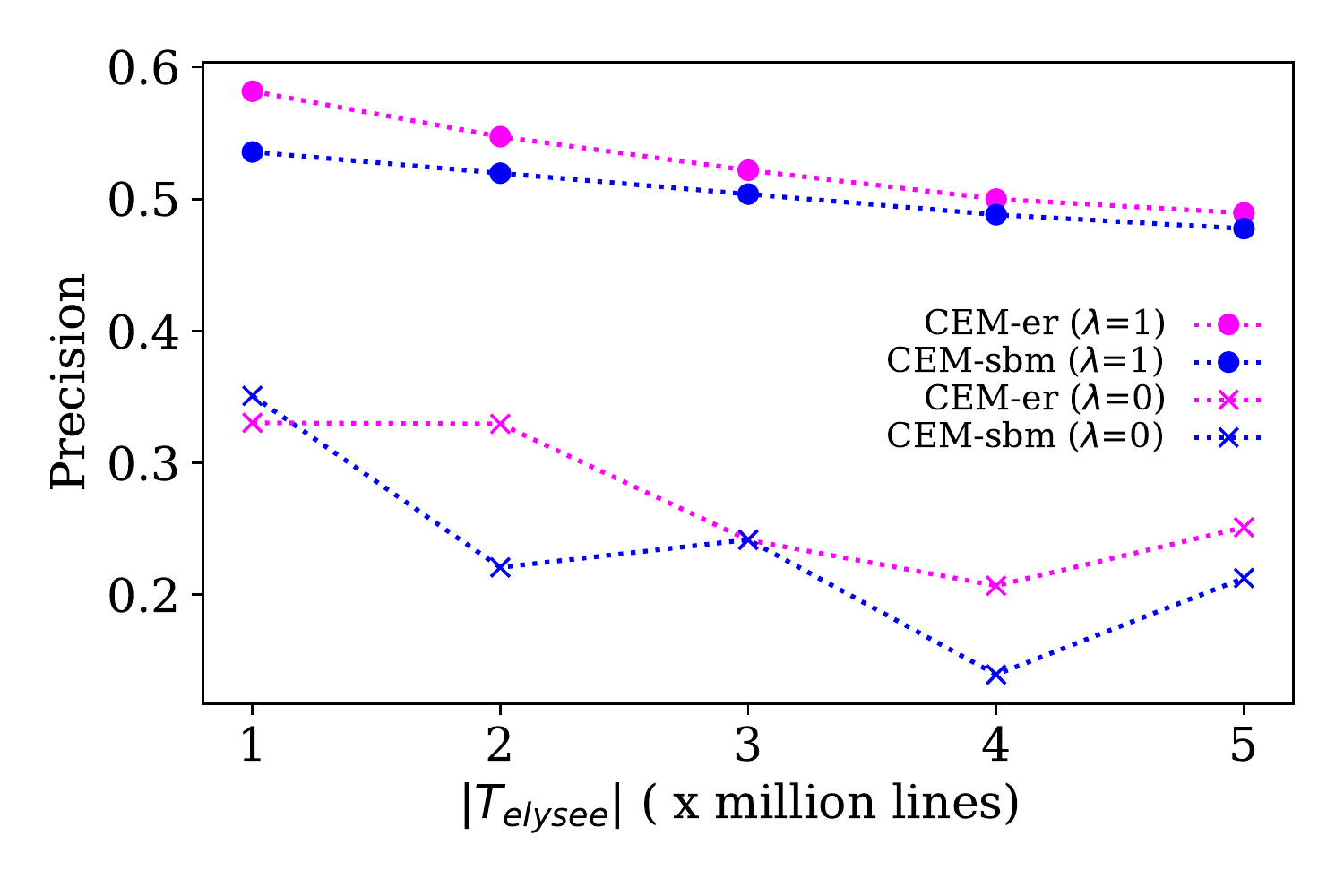}
		\end{subfigure}%
		\hfill
		\begin{subfigure}{.50\columnwidth}
			\centering
			\includegraphics[width=1\columnwidth]{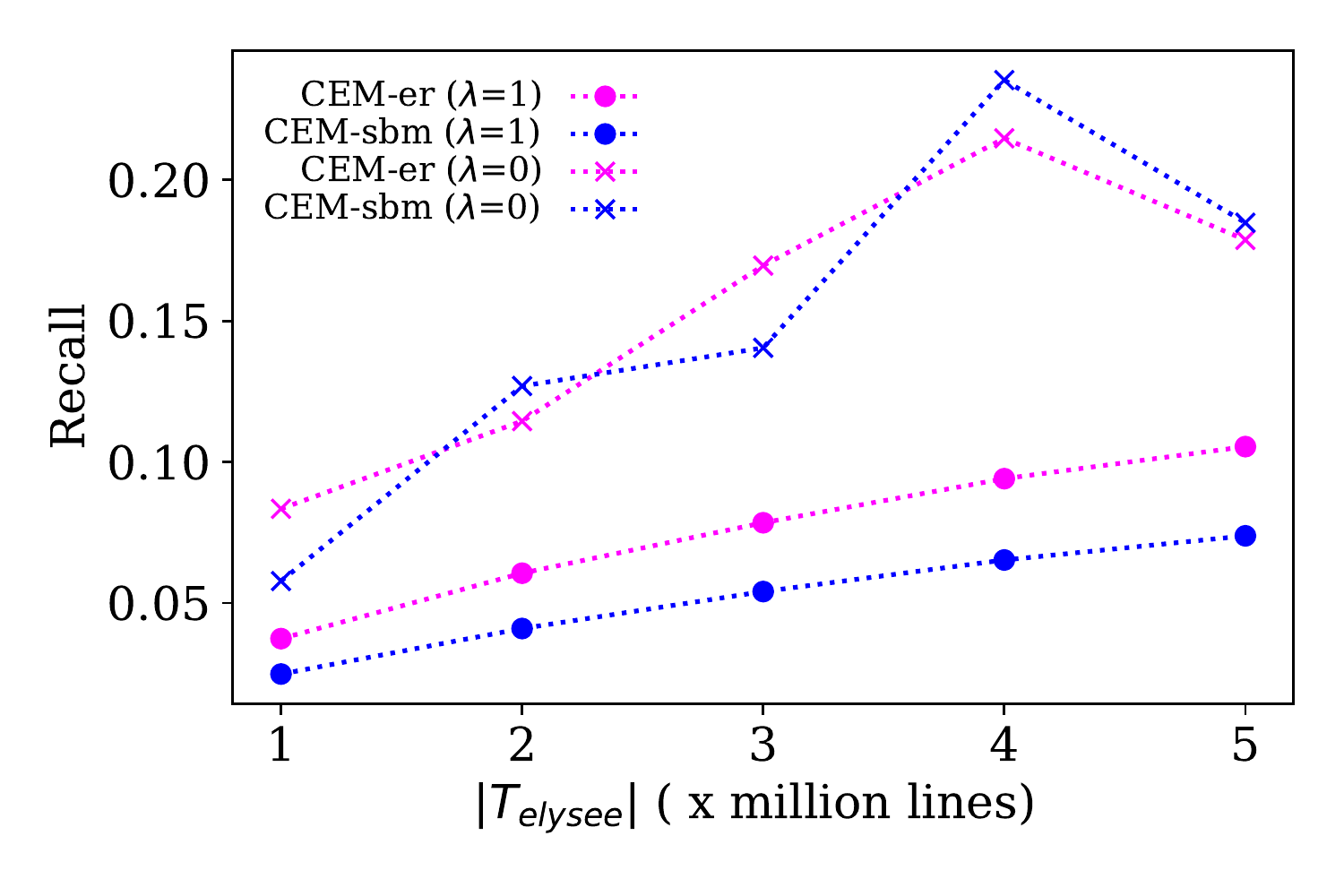}
			
		\end{subfigure}%
		\caption{For different sizes of the trace $T_{elysee}$ and $\lambda \in \{0,1\}$.}
		\label{fig:prandrecallelysee}
	\end{subfigure}
	\hfill
	\begin{subfigure}{\columnwidth}
		\begin{subfigure}{.50\columnwidth}
			\centering
			\includegraphics[width=1\columnwidth]{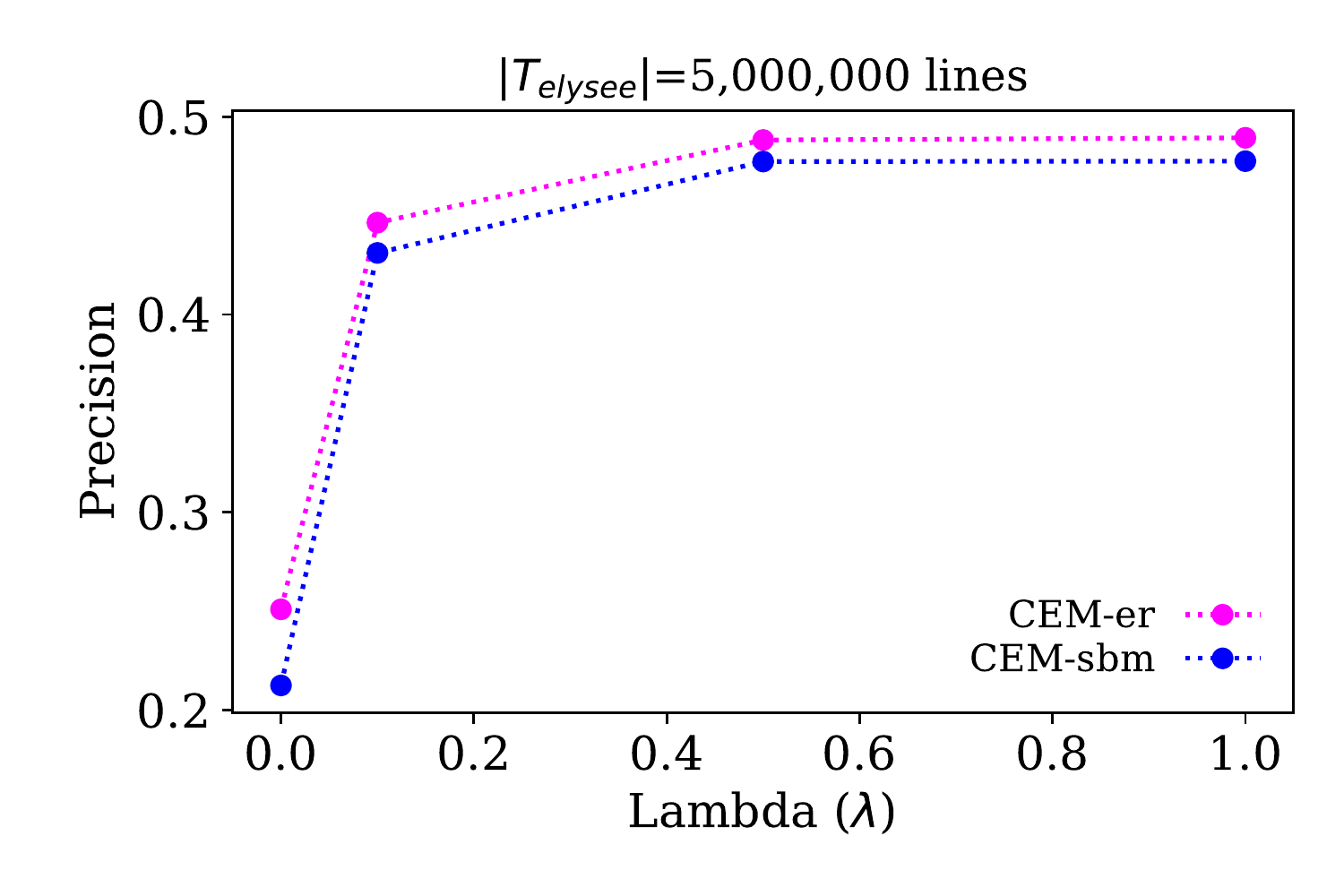}
			
		\end{subfigure}%
		\hfill
		\begin{subfigure}{.50\columnwidth}
			\centering
			\includegraphics[width=1\columnwidth]{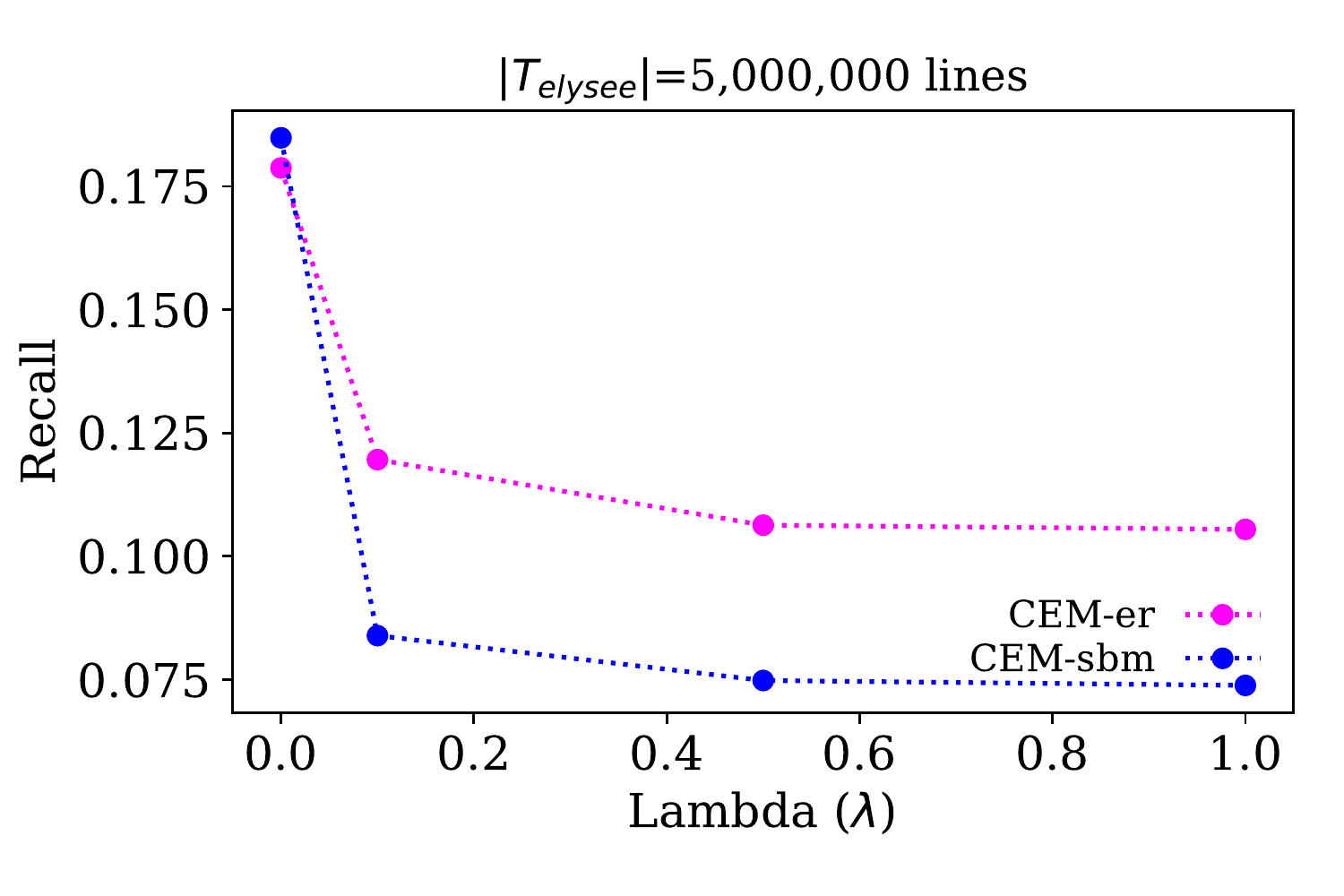}
			
		\end{subfigure}%
		\caption{For different values of $\lambda \in (0,1)$ given |$T_{elysee}$| = 5 mil. lines.}
		\label{fig:pr-rec-lambda-elysee}
	\end{subfigure}%
	\caption{Precision given Recall of CEM-er and CEM-sbm applied on \#Élysée2017fr.}
\end{figure*}

\subsection{Results of our method}

\textbf{Values of parameters.} The converged $\alpha, \beta$ parameters of CEM-er and CEM-sbm can be seen in Table \ref{tab:param-elysee}. Since all of the $\alpha$ values are close to 1, there is an almost 100\% probability that a post spread through an edge that we predicted to exist in the inferred networks. For CEM-er, the smaller value of $\beta$, which is almost equal to zero, suggests that there are zero false positive utilized edges. However, in the case of CEM-sbm ($\lambda=1$), the slightly higher value of $\beta^*=0.001$ suggests that there is a low, but existing probability, that a tweet passes via an edge that does not appear in the inferred ground truth. As we will see later, this may mean that we have missed some edges and therefore the overall feasibility rate may be (slightly) affected. Likewise, in the same case, the fact that $1-a^*=0.004$ means that there is a small probability of false negative utilized edges existing.

\textbf{Different sizes of input.} Figure \ref{fig:prandrecallelysee} shows the relation of Precision and Recall given trace sizes that range from 1 to 5 million lines. Again, as was the case with the synthetic data, the more information we have available, the higher the value of Recall will be. These values, however, will still stay at relatively low levels, under 0.1. As seen in Table \ref{tab:statistics}, this is largely due to the fact that only 45.23\% of the positive $(i,j)$ edges in the ground truth appear in the trace (i.e., they have $M_{ij}> 0$). The rest of them do not appear in the measurements, therefore it is not possible to infer them given the specific trace we have at hand. Still, we manage to predict thousands of edges that are mostly true positive (as seen from the Precision value). More specifically about Precision, we notice a slight drop as the size of the trace increases. This makes sense, since we infer more edges the more data we get, and therefore we are more likely to make errors. The drop is milder when $\lambda=1$ and more noticeable when $\lambda=0$.

\textbf{Different values of the hyperparameter $\boldsymbol \lambda$}. From Fig. \ref{fig:pr-rec-lambda-elysee} we notice that high values of $\lambda$ given a constant trace size (= 5 million lines) correspond to higher values of Precision. Here, we observe a trade-off between Precision and Recall, which was not evident in the synthetic dataset: in CEM-sbm for example, the lowest Precision($= 0.213$) corresponds to the highest Recall value($= 0.185$) when $\lambda=0$ and a lower Recall value ($=0.074$) corresponds to a higher Precision ($=0.478$) when $\lambda$ is set to 1. Therefore, we see that depending on our goal, we can choose to prioritize Precision over Recall and vice-versa. This can be controlled by the correct selection of the hyperparameter $\lambda$.

\textbf{Difference between priors.} In contrast to the synthetic dataset case, from the above figures we notice that CEM-er and CEM-sbm present more similar behavior. This is largely due to the properties of the trace itself: we have relatively sparse information on the edges between users that belong to different communities (we observe only 18.95\% of the existing inter-edges as seen in Table \ref{tab:statistics}, in contrast to the 98.25\% of the positive inter-edges in the case of the synthetic dataset). This makes sense since, in reality, users between different communities interact less often, so it is less likely that they will appear in a trace when we collect it. Therefore, the benefit of using the SBM instead of the ER prior cannot be easily made obvious given the specific trace that we have at hand. Still, the use of the SBM prior provides the highest Recall value ($=0.185$, for $\lambda=0$) and AUC value, ($=0.589$, for $\lambda=0$), which, as we will show later are also the largest values among all compared methods.

\subsection{Comparison between methods}

We compare the graphs inferred by our two models with the same methods presented before, this time when real-world data is given as input. Given our computational resources, we were not able to run the method by Peixoto (\citeyear{b12}) and Netinf (\citeyear{b2d}) within reasonable timeframes (in < 48 hours), therefore they are left out of the comparison. Table \ref{tab:elysee-perb} shows the Precision, Recall, and AUC performance of each method, and Table \ref{tab:tab8} shows the properties of each corresponding graph\footnote{N/A in the Tables refers to results not being available after 48 hours.}.

\begin{table*}[!t]
	\centering
	\caption{Performance of each method for the \#Élysée2017fr dataset.} 	
	\resizebox{0.60\linewidth}{!}{
		
		\begin{tabular}{llllll}
			\hline
			\addlinespace[0.1cm]
			{Performance} & {Precision} &  {Recall}  &  {AUC}  & {runtime(secs)}\\ 
			\hline
			\addlinespace[0.1cm]
			{Star}  & 0.446 & \text{0.133} & \text{0.565} & \textbf{1} 
			
			\\
			
			{Chain} & 0.262 & 0.130 & 0.563 & \textbf{1}  \\
			
			{Saito et al. (\citeyear{b5})} & 0.199 & 0.0001 & 0.500 & 342.00 \\
			
			{Netinf (\citeyear{b2d})} & N/A & N/A & N/A & N/A  \\ 		
			{Newman (\citeyear{b7})} & 0.464 $\pm$ 0.031 & 0.066 $\pm$ 0.001 & 0.533 $\pm$ 0.001  & \underline{25.00} \\
			
			{Peixoto (\citeyear{b12})} & N/A & N/A & N/A & N/A  \\
			\addlinespace[0.1cm] 
			\hline
			\addlinespace[0.1cm] 
			{CEM-er ($\lambda=0$)}  &  0.251 & \textbf{0.179} & \underline{0.586} &  37,721.00  \\
			{CEM-er ($\lambda=1$)}  & \underline{0.489} & 0.105 & 0.552 & 35,552.00  \\
			\addlinespace[0.1cm]								
			{CEM-sbm ($\lambda=0$)} &  0.213 & \underline{0.185} & \textbf{0.589} & 44,016.00 \\
			
			{CEM-sbm ($\lambda=1$)} &  \textbf{0.478} & 0.074 & 0.537 & 88,504.00 \\			
			\hline
			\addlinespace[0.1cm]
	\end{tabular}}
	\quad		
	\label{tab:elysee-perb}
\end{table*}

\begin{table*}[htbp]
	\centering
	\caption{Network statistics of the graphs inferred by each method compared to the ground truth graph for |$T_{elysee}$| = 5,000,000 lines.}
	\resizebox{\linewidth}{!}{%
		\begin{tabular}{lllllllllll}
			\hline
			\addlinespace[0.1cm]
			\text{Inferred network metrics}  &  {feasibility(\%)}  &  {\#edges}  &  {avg out-degree}  &  {max out-degree}  &  {max in-degree}  &  {diameter}  & {avg shortest path}  &  {max scc (\% users)} \\ \hline
			\addlinespace[0.1cm]
			Ground-truth  &  49.00 &  1,555,718 & 136.42 & 5,004 & 1,853 & 11 & 2.82 & 93.28 (10,747) 
			\\
			
			\hline
			\addlinespace[0.1cm]

			\text{Star}  &  \textbf{100.00}  & 463,290 & 40.25 & \textbf{2,524} & 1,069 & \textbf{12} & 3.70 & 66.05 (7,610) \\
			
			\text{Chain}  &  \textbf{100.00}  & 768,122 & 66.73 & 1,122 & 1,256 & 8 & \underline{3.04} & \textbf{95.34 (10,984)} 
			\\
			
			\text{Saito et al. (\citeyear{b5})}  & 0.55 & 786 & 0.54 & 2 & 10 & 8 & 1.11 & 0  \\
			
			{Netinf (\citeyear{b2d})} & N/A & N/A & N/A & N/A & N/A & N/A & N/A & N/A \\ 		
			\text{Newman (\citeyear{b7})} & 37.24  & 237,063  & 22.43  & 1,206 $\pm$ 127 & 558  & 12.7 & 4.32 & 52.57 (6,057)  
			%\text{Newman (\citeyear{b7})} & 37.24 $\pm$ 6.06 & 237,063 $\pm$ 12,369 & 22.43 $\pm$ 0.69 & 1,206 $\pm$ 127 & 558 $\pm$ 22 & 12.7 $\pm$ 0.68 & 4.32 $\pm$ 0.13 & 6,057 $\pm$ 351 
			\\
			
			\addlinespace[0.05cm]
			{Peixoto (\citeyear{b12})} & N/A & N/A & N/A & N/A & N/A & N/A & N/A & N/A \\	
			\addlinespace[0.1cm]
			\hline	
			\addlinespace[0.1cm]
			{CEM-er ($\lambda=0$)} & \textbf{100.00}  & \underline{1,108,079} & \underline{96.26} & \underline{2,336} & \underline{1,262} & \text{9} & 3.06 & 80.31 (9,252)
			
			\\
			
			{CEM-er ($\lambda=1$)} & \textbf{100.00}  & 335,289 & 29.13 & 2,291 & 790 & \underline{12} & 3.82 & 66.07 (7,612) 
			\\
			{CEM-sbm ($\lambda=0$)} & \textbf{100.00} & \textbf{1,353,432} & \textbf{117.58} & 1,364 & \textbf{1,609} & 8 & \textbf{2.95} & \underline{82.50 (9,505)} 
			\\
			{CEM-sbm ($\lambda=1$)} & \underline{99.37}  & 240,893 & 20.97 & 955 & 775 & \textbf{11} & 3.58 & 72.81 (8,388)
			\\

			\hline
	\end{tabular}}
	\label{tab:tab8}
\end{table*}

\subsubsection{Performance comparison}

From Tables	\ref{tab:elysee-perb} and \ref{tab:tab8} we observe that Star and Chain give 100\% feasible solutions with Precision equal to $0.446$ and $0.262$ respectively and Recall values equal to $0.133$ and $0.130$. However, their graph statistics resemble less these of the real graph: \textbf{Star} infers 463,290 edges, with max out-degree equal to $2,524$ and max in-degree equal to $1,069$. We consider these values quite high, given the number of edges inferred (they are comparable to the ground truth which has three times the number of edges of Star) and that's why we consider it less trustworthy. This result is expected due to the heuristic method of inferring the edges, which connects directly the author of a post to its reposters.

The graph by \textbf{Chain}, as seen in the last column of Table \ref{tab:tab8}, has the highest maximum strongly connected component (it includes $95.34\%$ of all users), which is bigger than the corresponding size in the real graph ($=93.28\%$). Given that the inferred graph by Chain is half the size of the real graph, this high percentage suggests that it is more densely connected than we would expect from a real graph. What is more, in a real-world graph, most nodes have a relatively small degree, but some of them will have a noticeably larger degree, being connected to many other nodes. However, in Chain, we do not notice this phenomenon.

%the maximum out and in-degrees are relatively small (=$1,122$ and $1,256$ respectively) given the size of its graph.

As was the case in the synthetic dataset evaluation, the method of Saito et al. (\citeyear{b5}) generates only a few edges ($=768$) and is therefore not feasible. The method of \textbf{Saito et al. (\citeyear{b5})} may be again relatively precise, but presents no strongly connected component, has a very low average out-degree ($=0.5$) and an abnormally high diameter ($=8$), given the size of the graph. The above shows that the graph inferred by this method is very sparse and does not resemble the real-world graph in question.

Likewise, the model by \textbf{Newman (\citeyear{b7})} is not feasible, but in this case, seems more competitive in terms of the Precision metric ($=0.464)$. However, its large diameter ($=12.7$) given the size of the inferred graph (5 times smaller than the real graph which has a diameter $=11$) prevents us from selecting it as a realistic option.

Compared with the above methods, our algorithm CEM-*, presents the highest values in terms of every metric: Precision, Recall, or AUC. This can be regulated either by choosing a value close to $\lambda=0$, that returns the highest number of nodes (>1,100,000) and therefore a high Recall (=$0.178$), for CEM-sbm ($\lambda=0$)) but lower Precision, or by choosing a value closer to $\lambda=1$ that returns less than 340,000 nodes (for both priors) and therefore a lower Recall but a high Precision (=$0.489$, CEM-er ($\lambda=1$)). When it comes to the statistics of the graph, its diameter stays close to the real value ($=11$). The same is true for the average shortest path. This illustrates that the two best values from each category are in favor of our CEM-* method.

\textbf{Optimization runtime.} We verify from the runtime column of Table \ref{tab:elysee-perb} that our model is scalable since we manage to solve an optimization problem with $6,922,990$ unknowns and $1,605,059$ constraints in only a couple of hours. We achieve this not only by formulating the inference as a linear optimization problem but also by taking advantage of powerful optimization solvers that are publicly available (in our case, the Gurobi solver). On the other hand, the methods by Saito et al. and Newman present fast computation times (342 and 25 seconds) but, as we have shown, they present less competitive results in terms of feasibility or performance.

\begin{figure}[]
	\centering
	
	\includegraphics[width=0.8\columnwidth]%
	{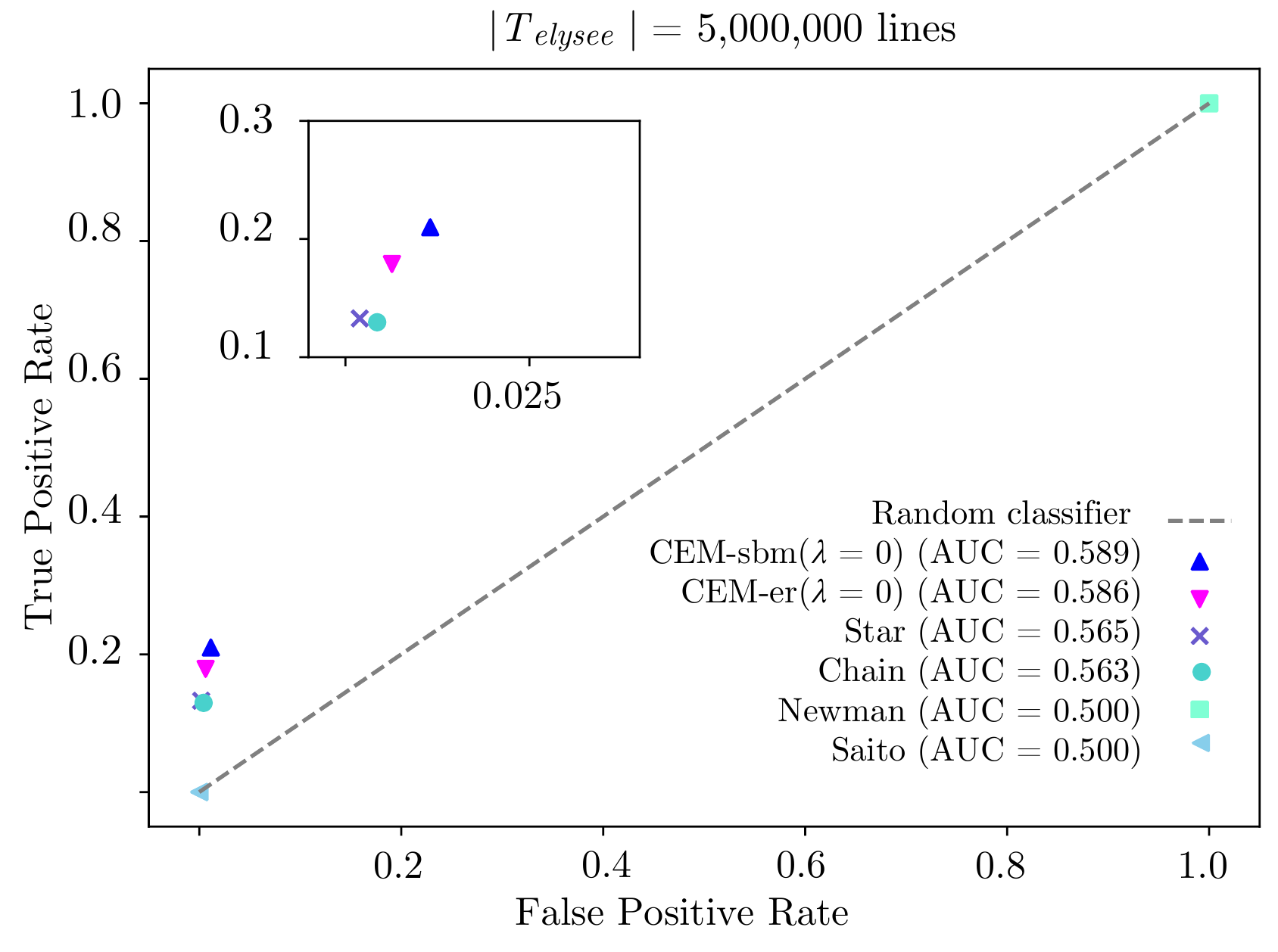}% picture filename
	\caption{True Positive related to the False Positive rates of each inference model when applied on \#Élysée2017fr.}
	\label{roc-elysee}
	\vspace{-0.3cm}	
\end{figure}

\textbf{ROC curve points of each method}. Again, on the upper left corner of the True Positive vs False Positive Rate figure (Figure \ref{roc-elysee}), we find our methods CEM-sbm ($\lambda=0$) and CEM-er ($\lambda=0$). Star and Chain are a bit lower, and the other methods are further away. This suggests that our model has the highest capacity to differentiate between the two classes (existing and non-existing edges) among all the other methods (and is also why we have the highest AUC values, as seen in Table \ref{tab:elysee-perb}).

	\begin{table*}[t]
	\caption{ Performance of community detection for the real-world graph with |$T_{elysee}|$ = 5,000,000 lines.}
	\hspace{1.6cm}
	\resizebox{0.5\columnwidth}{!}{
		\begin{tabular}{ll}
			\hline
			\addlinespace[0.1cm]
			\text{Label prediction} & 	{\text{F1-score}} \\ \hline
			\addlinespace[0.1cm]	
			{Star} & 0.858
			\\
			
			{Chain} & \textbf{0.889}
			\\
			
			{Saito et al. (\citeyear{b5})} & 0.447
			\\ 
			{Netinf (\citeyear{b2d})} & N/A \\ 
			{Newman (\citeyear{b7})} & \underline{0.888}  \\ 
			Peixoto (\citeyear{b12}) & N/A \\
			\addlinespace[0.1cm]
			\hline
			\addlinespace[0.1cm]
			
			{CEM-er ($\lambda=0$)}  & \underline{0.888} \\
			{CEM-sbm ($\lambda=0$)}  & 0.880   \\

			{CEM-er ($\lambda=1$)}  &  0.887 \\
			{CEM-sbm ($\lambda=1$)}  & 0.878\\

			\hline 
			\addlinespace[0.58cm]
	\end{tabular}}
	\resizebox{0.95\columnwidth}{!}{
		\begin{tabular}{lllll}
			\hline
			\addlinespace[0.1cm]
			{\text{Community parameters}} &  $p_{G}$ & |$\epsilon_{p}$| & $q_{G}$ & |$\epsilon_{q}$| \\ \hline
			\addlinespace[0.1cm]
			Ground-truth &  0.0012 & N/A & 0.0445 & N/A\\
			
			\hline
			\addlinespace[0.1cm]
			{Star} &  0.0143 & 10.92
			& {0.0003} & 0.99
			\\
			{Chain} & 0.0236 & 18.67 & {0.0004} & 0.99
			\\
			{Saito et al. (\citeyear{b5})} & 0.3834 &318.5 & N/A & N/A\\ 
			{Netinf (\citeyear{b2d}} & N/A & N/A&  N/A& N/A\\ 
			{Newman (\citeyear{b7}} &  \underline{0.0093} & \underline{6.75}
			& 5e-05 & 1
			
			\\ 
			Peixoto (\citeyear{b12} & N/A & N/A & N/A & N/A\\ 
			\addlinespace[0.1cm]
			\hline
			\addlinespace[0.1cm]
			{CEM-er ($\lambda=0$)}  &  0.0346 &27.83 & \underline{0.0005} & \underline{0.99} \\
			{CEM-sbm ($\lambda=0$)}  & 0.0425 & 34.42 & \textbf{0.0006} & \textbf{0.99} \\
			{CEM-er ($\lambda=1$)}  & 0.0103 & 7.58
			& 0.0002 & 1 \\
			{CEM-sbm ($\lambda=1$)}  & \textbf{0.0074} & \textbf{5.17}
			& 0.0001 & 1 \\
			%{CEM-sbm ($\lambda=1$) ($\beta$=0.7)}  &  \\
			
			\hline 
	\end{tabular}}
	
	\label{tab:communities:ely}
\end{table*}

\textbf{Detection of communities.} As shown in Table \ref{tab:communities:ely}, our methods CEM-er ($\lambda=0$) and CEM-er ($\lambda=1$) achieve a high F1-score ($=0.888$ and $0.887$), similarly to Newman's method ($=0.888$) and Chain ($=0.889$). Chain's high performance does not surprise us in this case since Chain favors the creation of communities all while inferring a very high number of edges compared to other methods. Despite this, all the $p$ parameters estimated on the graphs by each method are far from the real ground truth value. This was expected since we are missing substantial information on how edges interact between different communities and we may therefore be overestimating the value of $p$ while underestimating $q$. Still, our method for $\lambda=1$ has the lowest relative error on the $p$ parameter ($7.58$ for CEM-er and $5.17$ for CEM-sbm) along with Newman that has an $\epsilon_{p}=6.75$.

\begin{table*}[!t]
	\centering
	\caption{Performance of CEM-* given constant values of parameter $\beta$ for \#Élysée2017fr.}	
	\resizebox{0.55\linewidth}{!}{
		
		\begin{tabular}{llllll}
			\hline
			\addlinespace[0.1cm]
			{Performance given $\beta$} & {Precision} &  {Recall}  &  {AUC} & feasibility(\%) \\ 
			\hline
			\addlinespace[0.1cm]
			
			{CEM-er ($\lambda=1$)}  & 0.489 & \textbf{0.105} & \textbf{0.552} &  100.0  \\
			{CEM-er ($\lambda=1$) ($\beta=0.5$)} & \text{0.592}  & 0.060  & 0.530 &  66.96 \\
			{CEM-er ($\lambda=1$) ($\beta=0.6$)}  & \underline{0.604} & 0.052  & 0.526 &  61.21	 \\ 
			{CEM-er ($\lambda=1$) ($\beta=0.7$)}  & \textbf{0.619} & 0.040 & 0.520 &  \textbf{52.78} \\									
			\addlinespace[0.1cm]								
			{CEM-sbm ($\lambda=1$)} &  0.478 & \underline{0.074} & \underline{0.537} &   99.37\\
			{CEM-sbm ($\lambda=1$) ($\beta=0.5$)} &  0.552 & 0.054 & 0.527 &  71.86  \\			
			{CEM-sbm ($\lambda=1$) ($\beta=0.6$)} &  0.558 & 0.048 & 0.524 &  65.65 \\	
			{CEM-sbm ($\lambda=1$) ($\beta=0.7$)} &  0.566 & 0.041 & 0.520 &  \underline{58.00}\\					
			\hline
			\addlinespace[0.1cm]
	\end{tabular}}
	\quad		
	\label{tab:tab12}
\end{table*}

\subsection{Controlling feasibility through $\boldsymbol \beta$}

As expected, since 2017 (the year that the dataset was created), some Twitter profiles have been deleted or set to private. In addition, users may have retweeted a tweet/episode outside the scope of their followees (e.g., through Twitter search, recommendation algorithms, Twitter trends, etc.). As a result, the \#Élysée2017fr trace is not 100\% feasible given the ground truth friendship graph. In other words, the current view of the friendship graph does not explain all the episodes in the selected trace; in fact, it can only explain 49\% of them.  

We can therefore control the feasibility of our result to match the feasibility of the trace given the ground truth through the parameter $\beta$: for an inferred graph to be feasible, we want the false positive utilization rate $\beta$, i.e. the average number of inferred edges that pass through an edge that does not exist in the inferred graph, to be as close to 0 as possible. If $\beta$ is close to a non-zero value, it means that there is a $\beta>0$ probability that influence has happened through a nonexistent edge in the inferred graph and therefore some episodes may be left unexplained. Consequently, we can set $\beta$ equal to a constant - instead of updating it through Eq. \ref{ab} or \ref{eq41b} - whose value depends on the feasibility that we wish the outcome to have. Hence, we will examine the relation of the inferred graph to the ground truth given different constant values of $\beta$. In general, we expect the inferred graph to be more precise when the feasibility rate is close to this of the ground truth graph ($=$ 49\%).

First of all, as we show in Table \ref{tab:tab12} when $\beta$ increases feasibility decreases. For example, when $\beta=0.7$ the feasibility of the trace given the inferred graph is 52.78\% and 58\% for CEM-er and CEM-sbm respectively. We note that the value of $\beta$ changes only the overall number of edges inferred, which indirectly affects the number of episodes that are explained in the trace. Furthermore, as $\beta$ increases, and hence feasibility decreases, we get closer to the actual 50\% feasibility and precision improves. We should underline that when the feasibility rate falls lower than 50\% (for $\beta > 0.7 $), Precision falls dramatically since the algorithm starts inferring edges randomly, without really respecting the constraints.

\section{Evaluation with no ground truth } Overall, we notice that feasibility is proved beneficial and can increase the quality of the inferred graph. For example, the methods with the lowest feasibility rates (Saito et al., Netinf, Newman) infer graphs with low predictive quality and present statistics that are far from those of the real-world graph. The benefit of CEM-* over other methods is especially apparent when we have collected sufficient data between edges, as was the case in the synthetic dataset case. Consequently, when the underlying friendship graph is not available, which is often the case in graph reconstruction problems, the feasibility rate of the inferred graph could be an effective indicator of a method's performance. 

However, feasibility is not a sufficient condition for better prediction results. As we see in the case of Star, 
Chain, and CEM-* for $\lambda=0$, a 100\% feasibility rate cannot guarantee a precise result. Moreover, as we showed, we can use empirical values about how much feasibility to require in the inferred graph based, for example, on how old the trace is, or how often users retweet outside of their connections, e.g., using recommendations. On top of feasibility, we could look into the inferred graph's statistics and evaluate to what extent they are similar to these of a general, real-world graph. Usual indicators of such real-world properties are the average degree, the diameter, the average shortest path, and the strongly connected components of the graph.

\section{Conclusions}	

As we observed above, CEM-* successfully produces feasible graphs that are closer to reality when compared to heuristic and state-of-the-art methods. We validated the results both on synthetic and real-world traces, using two different graph priors, Erdős-Rényi (ER) and Stochastic Block Model (SBM), and noticed that CEM-* produces results that in most cases return the two most accurate values among all chosen compared metrics, and does so significantly faster than the state-of-the-art. Moreover, by selecting values between 0 and 1 for the hyperparameter $\lambda$, we can control the trade-off between the Precision and Recall of the result.

When comparing the effect of the two graph priors, we notice that the use of SBM can improve inference accuracy and community detection. The contribution of SBM is more apparent when we have sufficient information on how nodes interact between different communities.

Furthermore, we observe that feasible graphs are usually closer to the underlying graph compared to non-feasible graphs. When the trace is 100\% feasible given the ground truth, as is the case in the synthetic dataset, we find that feasibility is a necessary condition for the inferred graph to be as close to reality as possible. In the case of the \#Élysée2017fr graph, where the trace is only 50\% feasible, we observed that Precision improves as we force the feasibility of the inferred graph to be closer to the real percentage (through the parameter $\beta$). We conclude therefore that, for higher Precision, the feasibility of the trace given the inferred graph should match the feasibility of the trace given the true graph. However, if we cannot be sure about the ground truth's feasibility, we still suggest starting working with $\beta=0$, since, as we saw, it still returns better, and more realistic networks than other inference methods. Keep in mind, that as we saw in the case of Star and Chain, feasibility is not a sufficient condition for the accuracy of the result: the graphs inferred by both these methods are 100\% feasible but present some extreme properties (e.g., large diameter, low maximum degree) that make the results less trustworthy. 

We should note here that our method works with a specific trace structure that is based on the data that most social media platforms currently offer. In future work, we plan to apply our method and constraints to other types of data and graph inference cases to adapt to a wider range of domains (such as biology, epidemics, etc).

\begin{acknowledgements}
	
		An earlier version of this paper was presented at the 2021 IEEE/ACM International Conference on Advances in Social Networks Analysis and Mining 09-11 November 2021 (virtual) ASONAM 2021. This work is funded by the ANR (French National Agency of Research) by the “FairEngine” project under grant ANR-19-CE25-0011.
	
\end{acknowledgements}

\appendix

\section{CEM-er}
For the E-step, we modify the Newman algorithm by taking the expectation over the set of random variables $Y_{ij}$ at both sides of \eqref{eq7aa}: 
\begin{align}
\mathbb{E}[\log{P}(\theta \text{ }|\text{ } \Tc)]
\geq \mathbb{E} [\sum_{\textbf{A}} q(\textbf{A}) \log \frac{{P}(\textbf{A}, \theta\text{ }|\text{ } \Tc)}{q(\textbf{A}}] \nonumber \\
= \sum_{\textbf{A}} q(\textbf{A})\big(\mathbb{E}[\log {P}(\textbf{A}, \theta\text{ }|\text{ } \Tc)] - \log q(\textbf{A} \big)).
\label{eq7ba}
\end{align}
To find $\mathbb{E}[\log {P}(\textbf{A},\theta \text{ }|\text{ }\Tc)]$, we replace \eqref{eq8a} into \eqref{eq5a}. Setting $\Gamma={P}(\theta)/{P}(\Tc)$, the expectation of the log of \eqref{eq5a} becomes:
\begin{align}
\mathbb{E}[\log {P}(\textbf{A},\theta\text{ }|\text{ }\Tc)] = 
log \Gamma + \sum_{i \neq j} \Big[{A_{ij}} \Big(\log\rho + {\mathbb{E}[Y_{ij}]}\log\alpha + \nonumber \\
+(M_{ij}-\mathbb{E}[Y_{ij}])\log{(1 - \alpha)}\Big) 
\nonumber +(1-A_{ij}\Big(\log(1-\rho) + \nonumber\\
+ {\mathbb{E}[Y_{ij}]}\log\beta +(M_{ij} - \mathbb{E}[Y_{ij}])\log{(1 - \beta)\Big)\Big]}.
\label{eq7ca}
\end{align}
Then, by replacing \eqref{eq7d} into \eqref{eq7ca}, and then \eqref{eq7ca} into \eqref{eq7ba}, we get:
\begin{align} \label{eq7ferap}
\mathbb{E}[\log{P}(\theta \text{ }|\text{ } \Tc)] \geq \sum_{\textbf{A}} q(\textbf{A})\log\frac{D_{ij}}{q(\textbf{A})}
\end{align}

\begin{multline} \label{eq7fbaap}
\text{where, } D_{ij} = \Gamma \prod_{i \neq j}{\left[ \rho \alpha^{M_{ij}\sigma_{ij}}{(1 - \alpha)}^{M_{ij}(1-\sigma_{ij})}\right]}^{A_{ij}} \\
\times {\left[(1-\rho)\beta^{M_{ij}\sigma_{ij}}{(1 - \beta)}^{M_{ij}(1-\sigma_{ij})}\right]}^{1-A_{ij} }.
\end{multline}

For the M-step of the EM algorithm, the function that we want to maximize is $\mathbb{E}[\log{P}(\theta \text{ }|\text{ } \Tc)]$. To do so, we need to find the unknown values, $q(\textbf{A})$ and $\theta=$\{$\alpha, \beta, \rho, \boldsymbol{\sigma}$\}, that maximize the expectation on the left-hand side of \eqref{eq7fer}, under the feasibility constraints on the parameters set $\theta$. From these, only the $\sigma_{ij}$ have an important constraint set, specified in \eqref{eq4a} and \eqref{eq4}.

\textbf{Solution with respect to $\boldsymbol q(\textbf{A})$.} We notice that the choice of $q(\textbf{A})$ that achieves equality (i.e. maximizes the right-hand side) in \eqref{eq7fer} is: 
\begin{equation}
q(\textbf{A}) = \dfrac{D_{ij}}{\sum_{\textbf{A}}D_{ij}},
\label{eq9er}
\end{equation} 
which leads us to Eq. \eqref{eq22er}.

\textbf{Solution with respect to $\boldsymbol \alpha, \boldsymbol \beta, \boldsymbol \rho$.} To maximize the right-hand side of \eqref{eq7fer} in terms of parameter $\alpha$ we differentiate it with respect to $\alpha$ and then setting it equal to zero (while holding $\sigma_{ij}$, $q$ constant):
\begin{equation}
\sum_{i \neq j} Q_{ij} M_{ij} \left(\frac{\sigma_{ij}}{\alpha} - \frac{1-\sigma_{ij}}{1-\alpha}\right) = 0.
\label{eq18}
\end{equation}
After rearranging, we get the updates shown in Eq. \eqref{ab}, and we repeat likewise for $\beta$ and $\rho$.

\textbf{Solution with respect to $\boldsymbol \sigma_{ij}$.} If we take into account that $Q_{ij} = \sum_{\textbf{A}}q(\textbf{A})A_{ij}$ and also that $\sum_{\textbf{A}} q(\textbf{A}) = 1 $, by rearranging the right-hand side of \eqref{eq7fer}, the problem becomes equivalent to maximizing:

\begin{align}
\sum_{\textbf{A}} q(\textbf{A}) \sum_{i \neq j} \sigma_{ij} M_{ij}\left(A_{ij}\log\frac{\alpha}{1-\alpha} + (1-A_{ij})\log\frac{\beta}{1-\beta}\right) \ \nonumber 
\\ 
= \sum_{i \neq j} \sigma_{ij} M_{ij}\left(Q_{ij}\log\frac{\alpha}{1-\alpha}+ (1-Q_{ij})\log\frac{\beta}{1-\beta}\right).
\label{eq16}
\end{align}
This leads us to the constrained optimization problem of Eq. \eqref{eq17a}.

\section{CEM-sbm}

For the E-step of the EM algorithm, we modify the Newman algorithm by taking the expectation over the set of random variables $Y_{ij}$ at both sides of \eqref{eq7a}: 
\begin{align}
\mathbb{E}[\log{P}(\theta \text{ }| \text{ } \Tc)]
\geq \mathbb{E} [\sum_{\textbf{A}} q(\textbf{A}, \textbf{g}) \log \frac{{P}(\textbf{A}, \textbf{g}, \theta\text{ }|\text{ }\Tc)}{q(\textbf{A}, \textbf{g})}] \nonumber \\
= \sum_{\textbf{A}} q(\textbf{A}, \textbf{g})\big(\mathbb{E}[\log {P}(\textbf{A}, \textbf{g}, \theta\text{ }|\text{ }\Tc)] - \log q(\textbf{A}, \textbf{g} \big)).
\label{eq7b}
\end{align}
To find $\mathbb{E}[\log {P}(\textbf{A}, \textbf{g}, \theta \text{ }| \text{ } \Tc)]$, we replace \eqref{eq8} into \eqref{eq5}. Setting $\Gamma={P}(\theta)/{P}(\Tc)$, the expectation of the log of \eqref{eq5} becomes:

\begin{equation} 
\mathbb{E}[\log {P}(\textbf{A}, \textbf{g}, \theta \text{} \text{ }|\text{ } \text{} \Tc)] = 
log \Gamma + \sum_{\substack{{i \neq j}\\ {g_{i} = g{j}}}} \Big[{A_{ij}} \Big(\log p + {\mathbb{E}[Y_{ij}]}\log\alpha + \nonumber \\
\end{equation}

\begin{equation} 
+(M_{ij}-\mathbb{E}[Y_{ij}])\log{(1 - \alpha)}\Big) 
\nonumber +(1-A_{ij})\Big(\log(1-p) + \nonumber\\
\end{equation}

\begin{equation} 
+ {\mathbb{E}[Y_{ij}]}\log\beta +(M_{ij} - \mathbb{E}[Y_{ij}])\log{(1 - \beta)\Big)\Big]} + \sum_{\substack{{i \neq j}\\ {g_{i} \neq g{j}}}} \Big[{A_{ij}} \Big(\log q + \nonumber \\
\end{equation}

\begin{equation} 
+ {\mathbb{E}[Y_{ij}]}\log\alpha +(M_{ij}-\mathbb{E}[Y_{ij}])\log{(1 - \alpha)}\Big) 
\nonumber +(1-A_{ij})\Big(\log(1-q) + \nonumber
\end{equation}

\begin{equation} 
+ {\mathbb{E}[Y_{ij}]}\log\beta +(M_{ij} - \mathbb{E}[Y_{ij}])\log{(1 - \beta)\Big)\Big]}.
\label{eq7c}
\end{equation}
\normalsize By replacing \eqref{eq7d} into \eqref{eq7c}, and then \eqref{eq7c} into \eqref{eq7b}, we get:
\begin{align} \label{eq7f}
\mathbb{E}[\log{P}(\theta \text{ }|\text{ } \Tc)] \geq \sum_{\textbf{A}} q(\textbf{A}, \textbf{g})\log\frac{D(\textbf{A}, \textbf{g})}{q(\textbf{A}, \textbf{g})},
\end{align}
\normalsize
where, 
\begin{align}
{ D(\textbf{A}, \textbf{g})= \Gamma \prod_{\substack{{i \neq j}\\ {g_{i} = g{j}}}}{\left[ p \alpha^{M_{ij}\sigma_{ij}}{(1 - \alpha)}^{M_{ij}(1-\sigma_{ij})}\right]}^{A_{ij}}  }
\nonumber
\end{align}

\begin{multline} 
{\left[(1-p)\beta^{M_{ij}\sigma_{ij}}{(1 - \beta)}^{M_{ij}(1-\sigma_{ij})}\right]}^{1-A_{ij} } 
\prod_{\substack{{i \neq j} \\ {g_{i} \neq g{j}}}}{\left[ q \alpha^{M_{ij}\sigma_{ij}}{(1 - \alpha)}^{M_{ij}(1-\sigma_{ij})}\right]}^{A_{ij}} \nonumber
\end{multline}
\vspace{-0.5cm}
\begin{equation} \label{eq7fb}
{\left[(1-q)\beta^{M_{ij}\sigma_{ij}}{(1 - \beta)}^{M_{ij}(1-\sigma_{ij})}\right]}^{1-A_{ij} }.
\end{equation}
\normalsize

For the M-step of EM, we maximize the expectation $\mathbb{E}[\log{P}(\theta \text{ } |\text{ } \Tc)]$ as we did in the CEM-er prior. 

\textbf{Solution with respect to $\boldsymbol q(\textbf{A}, \textbf{g}$).} We notice that the choice of $q(\textbf{A}, \textbf{g})$ that achieves equality (i.e. maximizes the right-hand side) in \eqref{eq7f} is: 
\begin{equation}
q(\textbf{A}, \textbf{g}) = \dfrac{D(\textbf{A}, \textbf{g})}{\sum_{\textbf{A}}D(\textbf{A}, \textbf{g})}.
\label{eq9}
\end{equation} 
From \eqref{eq9}, in a similar fashion to Newman's method [Eq. (13), 20], and because $\Gamma$ cancels out, we get:
\begin{align}
q(\textbf{A}, \textbf{g}) = \prod_{i \neq j, (g_{i} = g_{j})}Q_{ij}(g_{i},g_{j})^{A_{ij}}(1-Q_{ij}(g_{i},g_{j}))^{1-A_{ij}} \nonumber
\\  \prod_{i \neq j, (g_{i} \neq g_{j})}Q_{ij}(g_{i},g_{j})^{A_{ij}}(1-Q_{ij}(g_{i},g_{j}))^{1-A_{ij}}.
\label{eq22}
\end{align}
Hence, given Eq. \eqref{eq7fb}, the values of $Q_{ij}$ are found to be the ones in Eq. \eqref{eq23sbm} and \eqref{eq23sbmb}.

Our goal is to find the unknown parameters $\theta=$\{$\alpha, \beta, p, q, \boldsymbol{\sigma}$\} that maximize the right-hand size of \eqref{eq7f}, given the maximising distribution for $q(\textbf{A}, \textbf{g})$ in \eqref{eq9}, hence given the values of $Q_{ij}(g_{i},g_{j})$ in \eqref{eq22}.

\textbf{Solution with respect to $\boldsymbol \alpha, \boldsymbol \beta, \boldsymbol p, \boldsymbol q$.}	To maximize the right-hand side of \eqref{eq7f} in terms of parameter $\alpha$, we differentiate the equation with respect to $\alpha$ and we set it equal to zero (while holding the rest of the parameters $\theta$ constant):
\begin{equation}
\sum_{i \neq j} Q_{ij}(g_{i},g_{j}) M_{ij} \left(\frac{\sigma_{ij}}{\alpha} - \frac{1-\sigma_{ij}}{1-\alpha}\right) = 0.
\label{eq18app}
\end{equation}
After rearranging, we get the value in Eq. \eqref{eq19b}. By repeating the same procedure for $\beta$, we get Eq. \eqref{eq41b}. Likewise, differentiating the r.h.s. of \eqref{eq7f} with respect to $p$ and then setting it equal to zero we get:
\begin{equation}
\sum_{\textbf{A}} q(\textbf{A, g}) \sum_ {\substack{{i \neq j}\\ {g_{i} = g{j}}}} (\frac{A_{ij}}{p} - \frac{1-A_{ij}}{1-p}) = 0.
\label{eq21bb}
\end{equation}
This is how we get the updates for $p$ in Eq. \eqref{eq21b}, and, likewise, for $q$ in Eq. \eqref{eq21c}.

\textbf{Solution with respect to $\boldsymbol \sigma_{ij}$.}  If we take into account that $Q_{ij}(g_{i},g_{j}) =\sum_{\textbf{A}}q(\textbf{A}, \textbf{g})A_{ij}$ and also that $\sum_{\textbf{A}} q(\textbf{A}, \textbf{g}) = 1 $, by rearranging the right-hand side of \eqref{eq7f}, the problem becomes equivalent to maximizing:

\begin{align}
\sum_{\textbf{A}} q(\textbf{A}, \textbf{g}) \sum_{i \neq j} \sigma_{ij} M_{ij}\left(A_{ij}\log\frac{\alpha}{1-\alpha} + (1-A_{ij})\log\frac{\beta}{1-\beta}\right) \ \nonumber 
\\ 
= \sum_{i \neq j} \sigma_{ij} M_{ij}\left(Q_{ij}(g_{i},g_{j})\log\frac{\alpha}{1-\alpha}+ (1-Q_{ij}(g_{i},g_{j}))\log\frac{\beta}{1-\beta}\right).
\label{eq16app}
\end{align}
This leads us to the optimization problem of Eq. \eqref{eq17asbm} through which we can find the $\sigma_{ij}$ values.

\end{document}